\documentclass[preprint,11pt]{JHEP3}

\usepackage{ifpdf}

\JHEPspecialurl{http://jhep.sissa.it/JOURNAL/JHEP3.tar.gz}

\usepackage{epsfig,multicol,amsmath}

\usepackage{epstopdf}
\usepackage{bm}  
\usepackage{amsmath}     
\usepackage{epsfig,epsf}
\usepackage{rotating}
\usepackage{cite}

\def\beq{\begin{equation}}
\def\eeq{\end{equation}}
\def\bea{\begin{eqnarray}}
\def\eea{\end{eqnarray}}

\def\eq#1{{Eq.~(\ref{#1})}}
\def\fig#1{{Fig.~\ref{#1}}}
\newcommand{\bas}{\bar{\alpha}_S}
\newcommand{\as}{\alpha_S}

\newcommand{\Lb}{\left(}
\newcommand{\Rb}{\right)}
\parskip=\itemsep               

\newcommand{\nn}{\nonumber}

\newcommand{\h}{\frac{1}{2}}

\newcommand{\pom}{I\!\!P}

\def\pom{{I\!\!P}}
\def\reg{{I\!\!R}}
\title{
A CGC/saturation approach for angular correlations
  in proton-proton scattering. }
\author{\Large  E. Gotsman$^{a}$\thanks{Email:
gotsman@post.tau.ac.il.} , \,\, E. Levin$^{a,b}$\thanks{Email:
leving@post.tau.ac.il} \,\,and I. Potashnikova$^{b}$\thanks{Email:irina.potashnikova@usm.cl}
\\
a)\,  \,Department of Particle Physics, School of Physics and Astronomy,
Raymond and Beverly Sackler
 Faculty of Exact Science, Tel Aviv University, Tel Aviv, 69978, Israel\\ 
b)\,\,Departamento de F\'\i sica, Universidad T\'ecnica Federico Santa 
Mar\'\i a, Avda. Espa\~na 1680\\ and Centro 
Cientifico-Tecnol$\acute{o}$gico de Valpara\'\i so,Casilla 110-V, 
Valparaiso, Chile

\\}

\abstract{ We generalized our model for the description of 
 hard processes, and  calculate  the value  of the  azimuthal angle
 correlations ( Fourier
 harmonics $v_n$ ), for proton-proton scattering.
The energy and 
 multiplicity independence, as well as the value of $v_n$, turn out
 the be in accord with the experimental data, or  slightly larger.
  Therefore,  before making extreme assumptions on proton-proton
 collisions,  such as the production of quark-gluon plasma in the 
large
 multiplicity events, we need to explain  how these effect the
 Bose-Einstein correlations which are so large, that have to be
 taken into account,      and which are able to describe the
 angular correlations in proton-proton collisions, without including
 final state interactions. }

\dedicated{PACS: 13.85.-t, 13.85.Hd, 11.55.-m, 11.55.Bq}
\preprint{TAUP - 3008/16\\
{\tt }
\today}
\begin{document}

\section{Introduction}

The experimental data on azimuthal angle correlations (see Refs.
\cite{CMSPP,STARAA,PHOBOSAA,STARAA1,CMSPA,CMSAA,ALICEAA,ALICEPA,
ATLASPP,ATLASPA,ATLASAA} show  suprizing similarities between 
different
 processes: nucleus-nucleus, hadron-nucleus and hadron-hadron collisions.
 The  popular explanation is related to  elliptic flow, and stems
 from the interaction in the final state. In the framework of such an 
approach,
 we have to assume that the proton interactions  are similar to 
nucleus
 scattering, at least for events with large multiplicity. However,
 the ATLAS data \cite{ATLASPP} show that $v_{2,2}$,$v_{3,3}$ and $v_{4,4}$
 do not depend on multiplicity at W=13 TeV and at W= 2.76 TeV. 

In this paper we will discuss these correlations from a different point
 of view. We believe that the general origin of the azimuthal angle
 correlations in all reactions, stems from the Bose-Einstein 
 correlations (BEC) of the produced gluons, which   originate from  
 the
  gluon wave function in the initial state\cite{
PION,GLMBEH,KOWE,KOLU1,KOLUCOR,GOLE,GLMBE,KOLULAST,RAJUREV,KOLUREV}. The 
 attractive feature of this
 idea is that, BEC have a general source that characterizes the 
volume
 of the interaction\cite{HBT,IPCOR}. Therefore, the main 
dimensional parameters of the interaction  that manifest themselves
 in diffraction scattering, and in inclusive production, should determine
 the BEC. In other words, that in spite of the embryonic stage of our
 understanding of the confinement of quarks and gluons, we can develop
a quantitative approach for the BEC in the framework of a  model for
 soft interactions at high energy.
 To accomplish this, we need to  construct such a model, that 
will allow us to discuss
  soft and  hard processes on the same footing.
 
 The main goal of this paper is to develop such model. Fortunately,
 we have built  a model which  provides a good description of all 
the soft
 data\cite{GLMNI,GLM2CH,GLMINCL,GLMCOR,GLMSP,GLMACOR}, including, total,
 inelastic, elastic and diffractive cross sections, the $t$-dependence of
 these cross sections as well as the inclusive production and rapidity
 correlations. In this paper we expand this model to include the 
hard interactions
  mostly using the geometric scaling behavior of the scattering
 amplitude\cite{GS}  for   the hard kinematic region in the Colour
 Glass Condensate (CGC)/saturation approach\cite{KOLEB}.
 
 The idea of  BEC  being the main  source of the azimuthal angle 
correlations,
 is marred by the observation\cite{KOWE,KOLUCOR}, that the process of the
 central diffraction production of  colourless gluon dijets, gives 
a
 contribution which is equal to that of  the BEC. In this case $v_{n,n}$ 
with
 odd $n$ are equal to zero, while   $v_{n,n}$ with even $n$, are twice 
larger.
  We will not  discuss this problem here. Our main
 goal is to obtain reliable quantitative  estimates for $v_{n,n}$. 
However,
 we believe that due to the Sudakov suppression in Double Log 
Approximation of
 perturbative QCD,  the dijet contribution is negligibly small
 \cite{GOLELAST}.
 
 The paper is organized as follows. In the next section we  give a
 brief review of our model  which is based on CGC/saturation 
approach. We 
 discuss what we have taken from the theory in our approach, and what
 we have considered from a pure phenomenological approach. We will attempt 
to
 clarify the physical meaning of the introduced phenomenological 
parameters,
 and show how we include three dimensional sizes, which have been used to
 describe the scattering amplitude. In the language of the Constituent
 Quark Model these three sizes are the hadron radius, the size of the
 constituent quark, and the saturation momentum, 
 which is a typical scale for the high energy amplitude. 
 
 In section 3 we generalize our model including the 
 construction of an amplitude at short distances, which is able to  
describe
 the deep inelastic scattering (DIS). We compare our amplitude with HERA
 experimental data \cite{HERA}. In  section four we   calculate the
 inclusive cross section, and show that we  obtain good agreement 
 with the experimental data. This is very important for our calculation, 
since
  it demonstrates that we are able to describe the experimental data 
for 
inclusive production
 not only at short distances, but also at long distances. 
 
 In section 5 we calculate the value of  BEC correlations, its energy
 and multiplicity dependence. We 
 obtain the value of $v_n$ which are a bit larger than the experimental
 ones, with a mild dependence on energy and multiplicity. We consider 
these
 estimates as the first quantitative prediction for $v_n$ in proton-proton
 scattering, which are in agreement with the values of the inclusive cross
 sections, and the cross sections for the hard processes.
 
 In section 6 we  draw our conclusions and outline the problems for
 future investigations.
 
 \section{The model} 
 
 \subsection{Theoretical input from  the CGC/saturation approach}
 In this section we generalize our model for  soft
 interactions at high 
energy\cite{GLMNI,GLM2CH,GLMINCL,GLMCOR,GLMSP,GLMACOR} 
 to include   a description of  hard processes. This model incorporates two
 ingredients: the achievements of the CGC/saturation approach, which is
 an effective theory for  QCD at high energy; and the pure  
phenomenological
 treatment of the long distance non-perturbative  physics, due to the lack 
of
 the theoretical understanding of confinement of quark and gluons.
 
  We wish to stress that the most of this section does not contain
 new results, it reviews our approach, and it is included in the paper 
only
 for the sake of completeness of presentation.  One can find 
more details in
 Refs.\cite{GLMNI,GLM2CH,GLMINCL,GLMCOR,GLMSP,GLMACOR}.

   The effective theory for QCD at high energies
 exists in two different formulations:  the CGC/saturation approach
 \cite{MV,MUCD,BK,JIMWLK}, and the BFKL Pomeron calculus 
\cite{BFKL,LI,GLR,GLR1,MUQI,MUPA,BART,BRN,KOLE,LELU,LMP,AKLL,AKLL1,LEPP}. 
 In building our model we rely on the BFKL Pomeron calculus, as
 the relation to diffractive physics  and soft processes in general,
 is more transparent  in this approach.
  However, we believe  the CGC/saturation approach  produces a more 
general
  pattern \cite{AKLL,AKLL1}   for the treatment of high energy QCD.
  Fortunately, in Ref.\cite{AKLL1} it was shown, that 
these
 two approaches are equivalent for
\beq \label{MOD1}
Y \,\leq\,\frac{2}{\Delta_{\mbox{\tiny BFKL}}}\,\ln\Lb
 \frac{1}{\Delta^2_{\mbox
{\tiny BFKL}}}\Rb
\eeq
where $\Delta_{\mbox{\tiny BFKL}}$ denotes the intercept of the BFKL 
 Pomeron. As we will see, in our model $ \Delta_{\mbox{\tiny BFKL}}\,
\approx\,0.2 - 0.25$    leading to $Y_{max} = 20 - 30$, which covers
 all accessible energies.
 
\begin{figure}[ht]
    \centering
  \leavevmode
      \includegraphics[width=14cm]{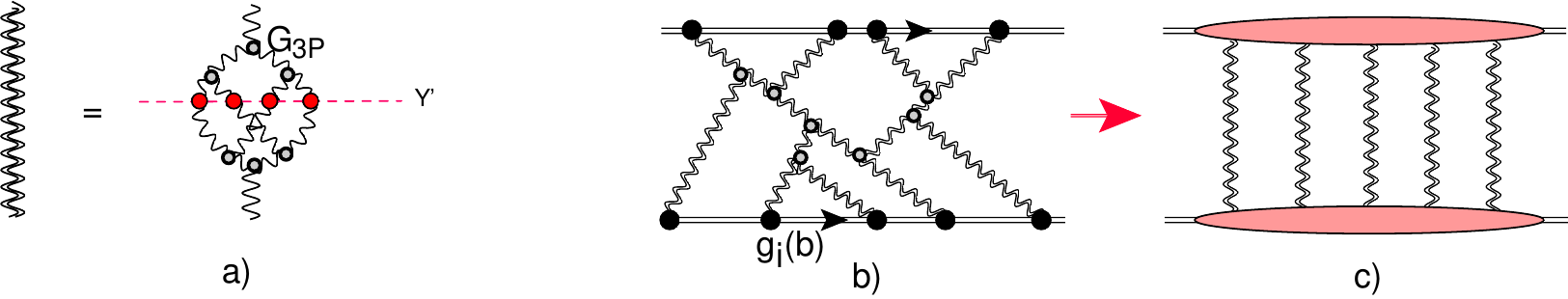}  
      \caption{\protect\fig{amp}-a shows the set of  diagrams in the
 BFKL Pomeron calculus that produce the resulting (dressed) Green
 function of the Pomeron in the framework of high energy QCD. The red blobs
 denote the amplitude for the dipole-dipole interaction at low energy.
 In \protect \fig{amp}-b the net diagrams,    which   include
 the interaction of the BFKL Pomerons with colliding hadrons, are shown.
 The sum of the diagrams after integration
 over positions of $G_{3 \pom}$ in rapidity, reduces to   \protect\fig{amp}-c.}
\label{amp}
   \end{figure}

  The main ingredient, that we need to find, is the resulting (dressed)
 BFKL Pomeron Green function, which 
 can be calculated using $t$-channel unitarity constraints:
\bea \label{GFMPSI}
 &&G^{\mbox{\tiny dressed}}_\pom\Lb Y, r, R; b \Rb\,\,=\\
 &&\,\,\int \prod_{i =1} d^2 r_i\, d^2 b_i\, d^2 r'_i\, d^2 b'_i\,
 N\Lb Y- Y', r, \{ r_i,b - b_i\}\Rb\fbox{$A^{\rm BA}_{\mbox{\tiny
 dipole-dipole}}
\Lb r_i, r'_i, \vec{b}_i - \vec{b'}_i\Rb$}\,
N\Lb  Y', R, \{ r'_i,b'_i\}\Rb\nn
\eea
where  $ N\Lb Y- Y', r, \{ r_i,b - b_i\}\Rb$ denotes the amplitude
  for the  production in the  $t$-channel  of  the set of dipoles 
with $Y=Y'$
 and with the size $r_i$,
at the impact parameters $b_i$.
$A^{BA}_{\mbox{\tiny dipole-dipole}}$ denotes the dipole-dipole
 scattering amplitude in the Born approximation of perturbative
 QCD, which are   indicated by red circles  in \fig{amp}-a.  
 In addition, in Ref.\cite{AKLL1} it is shown that  for such $Y$, we can
 safely use the Mueller-Patel-Salam-Iancu (MPSI) approach\cite{MPSI}.
 In this approximation we estimate the amplitudes $N$ in \eq{GFMPSI},
 using  BFKL Pomeron 'fan'  diagrams (see \fig{amp}-a for examples of
 such diagrams). In other words, we can use the parton cascade of the
 Baslitsky-Kovchegov\cite{BK} equation,  
 to find 
  the amplitude for the production of dipoles of size 
$r_i$ at
  impact parameters $b_i$.  This amplitude can be 
written as (see \fig{amp}-c)
    \bea \label{TI3}
&&N\Lb Y- Y', r, \{ r_i,b_i\}\Rb\,\,=\,\,N^{\rm BK}\Lb Y- Y', r, \{ r_i,b_i\}\Rb\\
&&=\,\,\sum^{\infty}_{n=1} \,\Lb - \,1\Rb^{n+1} \widetilde{C}_n\Lb
  r\Rb \prod^n_{i=1} G_\pom\Lb Y - Y';  r, r_i , b_i\Rb\,\,
=\,\,\sum^{\infty}_{n=1} \,\Lb - \,1\Rb^{n+1} \widetilde{C}_n\Lb
 r\Rb \prod^n_{i=1} G_\pom\Lb z - z_i\Rb\nn.
\eea   
$G_\pom$  denotes the Green function of the BFKL Pomeron. In the last
 equation we used the fact that in the saturation region this Green 
function
 has  geometric scaling behavior, and so it depends on one variable:
 $z_i \,=\,\ln\Lb Q^2_s(Y') r^2_i\Rb$, where $Q_s\Lb Y'\Rb$, is the
 saturation scale,  in the vicinity of the saturation scale\cite{MUTR}
\beq \label{VQS1}
G_\pom\Lb  z_i\Rb\,=\,\phi_0 \Lb r^2_i\,Q^2_s\Lb Y, b_i\Rb\Rb^{1 - \gamma_{cr}}
\eeq
where $\gamma_{cr}=0.37$.

  In Ref.\cite{LEPP}, it was shown that, 
 the solution to the non-linear BK equation has   the following general 
form
 \beq \label{TI4}
N\Lb G_\pom\Lb \phi_0,z\Rb\Rb \,\,=\,\,\sum^{\infty}_{n=1} \,\Lb - 
\,1\Rb^{n+1} C_n\Lb \phi_0\Rb G_\pom^n\Lb \phi_0,z\Rb.
\eeq   
Comparing \eq{TI3} with \eq{TI4} we see 
\beq \label{TI5}
\widetilde{C}_n\Lb  r\Rb\,\,\,=\,\,\,C_n\Lb \phi_0\Rb.
\eeq 
Coefficients $C_n$ can be determined from the solution to the 
Balitsky-Kovchegov
 equation \cite{BK}, in the saturation region. The numerical
 solution has been found in Ref.\cite{LEPP} for the simplified
 BFKL kernel in which only the leading twist contribution was taken into 
account:
 
\beq \label{T16}
N^{\rm BK}\Lb G_\pom\Lb \phi_0,z\Rb\Rb \,\,=\,\,a\,\Lb 1
 - \exp\Lb -  G_\pom\Lb \phi_0,z\Rb\Rb\Rb\,\,+\,\,\Lb 1 - a\Rb
\frac{ G_\pom\Lb \phi_0,z\Rb}{1\,+\, G_\pom\Lb \phi_0,z\Rb},
\eeq
with $a$ = 0.65. \eq{T16} is a convenient parameterization of the
 numerical solution, with an accuracy of better than 5\%.
Having $C_n$ we can calculate the Green function of the dressed BFKL
 Pomeron using \eq{GFMPSI}, and the property of the BFKL Pomeron exchange:
\bea \label{POMTUN}
&&\frac{\as^2}{4 \pi} \,\,G_\pom\Lb Y - 0, r, R; b  \Rb\,=\,\\
&&\int d^2 r'
  d^2 b' \, d^2 r'' \,d^2 b'' \,G_\pom\Lb Y - Y', r, r', \vec{b}
 - \vec{b}^{\,'} \Rb \, \,
\,G_\pom\Lb Y' r'', R,  \vec{b}^{\,''} \Rb\,\,A^{\rm BA}_{\mbox{\tiny
 dipole-dipole}}\Lb r', r'', \vec{b''} - \vec{b'}\Rb\nn
\eea 
 
Carrying out the integrations in \eq{GFMPSI}, we obtain the Green
 function of the dressed Pomeron in the following form:
 
 \bea \label{G}
G^{\mbox{\tiny dressed}}\Lb T\Rb\,\,&=&\,\,a^2 (1 - \exp\Lb -T\Rb )  +
 2 a (1 - a)\frac{T}{1 + T} + (1 - a)^2 G\Lb T\Rb \nn\\
~~~&\mbox{with}&~~G\Lb T\Rb = 1 - \frac{1}{T} \exp\Lb \frac{1}{T}\Rb
 \Gamma\Lb 0, \frac{1}{T}\Rb
\eea
where $\Gamma\Lb s, z\Rb$ is the upper incomplete gamma function
 (see Ref.\cite{RY} formula {\bf 8.35}), and $T$ denotes  the BFKL Pomeron 
in 
the
 vicinity of the saturation scale ( see \eq{VQS1})
  \beq \label{T}
T\Lb r_\bot, Y\,=\,\ln\Lb s/s_0\Rb, b\Rb\,\,=\,\,\phi_0  \Lb r^2_\bot Q^2_s\Lb Y, b\Rb\Rb^{\bar
 \gamma}  
\eeq
 
The Green function of \eq{G} depends on the size of the dipoles, and we
 will use it for discussing the hard processes. In our analysis of the
 soft interaction we fixed $r = 1/m$, and $m$ was a fitting parameter.

 \subsection{Phenomenology: assumptions and  new small parameters }

 
 Unfortunately, due to the embryonic stage of theoretical understanding 
of the confinement of quarks and gluons,  it is necessary to use pure
 phenomenalogical ideas to fix two major problems in high energy 
scattering: the structure of hadrons, and the large impact parameter 
behavior
 of the scattering amplitude\cite{KOWI}.
 The main idea  to  correct the large impact parameter
 behaviour,  is to assume that the saturation momentum  has the 
following dependence  on the 
impact parameter $b$: 
    \beq \label{QS}
Q^2_s\Lb b, Y\Rb\,\,=\,\,Q^2_{0s}\Lb b, Y_0\Rb\,e^{\lambda \,(Y - Y_0)}
\eeq
with
\beq \label{QS0}
Q^2_{0s}\Lb b, Y_0\Rb\,\,=\,\, \Lb m^2\Rb^{1 - 1/\bar \gamma}\,\Lb S\Lb b,
 m\Rb\Rb^{1/\bar{\gamma}} 
~~~~~~~S\Lb b , m \Rb \,\,=\,\,\frac{m^2}{2 \pi} e^{
 - m b}~~~\mbox{and}~~\bar \gamma\,=\,0.63
\eeq 

 We have introduced a new phenomenological parameter $m$ to
 describe the large $b$ behaviour. The $Y$ dependence  as well
 as  $r^2$ dependence, can be found from CGC/saturation approach 
\cite{KOLEB},
 since $\phi_0$ and $\lambda$ can be calculated in the leading order of
 perturbative QCD. However, since the higher order corrections turn out
 to be large \cite{HOCOR}, we treat them as parameters to be fitted. $m$
 is a non-perturbative parameter, which determines the typical sizes of
 dipoles within the hadrons. In Table 1, we show that from the fit,  
 $m$ = 5.25 GeV, supporting our main assumption that we can
 apply the BFKL Pomeron calculus, based on perturbative QCD, to the
 soft interaction since $m \,\gg\,\mu_{soft}$, where $\mu_{soft}$ is
 the scale of soft interaction, which is of the order of the mass of
 pion or $\Lambda_{\rm QCD}$.
 
 The idea to absorb the non-perturbative $b$ dependence into the saturation
 scale, stems both from the success of this idea in the description of the
 hard processes in framework of the saturation model
 \cite{IIM,SATMOD0,SATMOD1, SATMOD2,SATMOD3, SATMOD4, SATMOD5,
 SATMOD6,SATMOD7, SATMOD8, SATMOD9,SATMOD10,SATMOD11,SATMOD12, SATMOD13,
 SATMOD14,SATMOD15,SATMOD16, SATMOD17, CLP,CLP1}, and from the 
semi-classical
 solution to the BK equation\cite{BKL}, as well as from the analytical
 solution deep in the saturation domain\cite{LETU}.
 
 The second unsolved problem for which we need  a phenomenological input,
 is the structure of the scattering hadrons.
 We use a two channel model, which allows us to calculate the
 diffractive production in the region of small masses.
   In this model, we replace the rich structure of the 
 diffractively produced states, by a single  state with the wave 
function 
$\psi_D$, a la Good-Walker \cite{GW}.
  The observed physical 
hadronic and diffractive states are written in the form 
\beq \label{MF1}
\psi_h\,=\,\alpha\,\Psi_1+\beta\,\Psi_2\,;\,\,\,\,\,\,\,\,\,\,
\psi_D\,=\,-\beta\,\Psi_1+\alpha \,\Psi_2;~~~~~~~~~
\mbox{where}~~~~~~~ \alpha^2+\beta^2\,=\,1;
\eeq 

Functions $\psi_1$ and $\psi_2$  form a  
complete set of orthogonal
functions $\{ \psi_i \}$ which diagonalize the
interaction matrix $T$
\beq \label{GT1}
A^{i'k'}_{i,k}=<\psi_i\,\psi_k|\mathbf{T}|\psi_{i'}\,\psi_{k'}>=
A_{i,k}\,\delta_{i,i'}\,\delta_{k,k'}.
\eeq
The unitarity constraints take  the form
\beq \label{UNIT}
2\,\mbox{Im}\,A_{i,k}\left(s,b\right)=|A_{i,k}\left(s,b\right)|^2
+G^{in}_{i,k}(s,b),
\eeq
where $G^{in}_{i,k}$ denotes the contribution of all non 
diffractive inelastic processes,
i.e. it is the summed probability for these final states to be
produced in the scattering of a state $i$ off a state $k$. In \eq{UNIT} 
$\sqrt{s}=W$ denotes the energy of the colliding hadrons, and $b$ 
the 
impact  parameter.
A simple solution to \eq{UNIT} at high energies, has the eikonal form 
with an arbitrary opacity $\Omega_{ik}$, where the real 
part of the amplitude is much smaller than the imaginary part.
\beq \label{A}
A_{i,k}(s,b)=i \Lb 1 -\exp\Lb - \Omega_{i,k}(s,b)\Rb\Rb,
\eeq
\beq \label{GIN}
G^{in}_{i,k}(s,b)=1-\exp\Lb - 2\,\Omega_{i,k}(s,b)\Rb.
\eeq
\eq{GIN} implies that $P^S_{i,k}=\exp \Lb - 2\,\Omega_{i,k}(s,b) \Rb$, is 
the probability that the initial projectiles
$(i,k)$  reach the final state interaction unchanged, regardless of 
the initial state re-scatterings.
\par

 The first approach is to use the eikonal approximation for $\Omega$ in which
 \beq \label{EAPR}
 \Omega_{i,k}(r_\bot, Y - Y_0,b)\,\,=\,\int d^2 b'\,d^2 b''\,
 g_i\Lb \vec{b}',m_i\Rb \,G^{\mbox{\tiny dressed}}\Lb T\Lb r_\bot,
 Y - Y_0, \vec{b}''\Rb\Rb\,g_k\Lb \vec{b} - \vec{b}'\ - \vec{b}'',m_k\Rb 
 \eeq 
 where $m_i$ denote the masses, which is introduced phenomenologically to
 determine the $b$ dependence of $g_i$ (see below).
 
 We propose a more general approach, which takes into account the new
 small parameters, that are determined by fitting to the experimental 
data
 (see Table 1 and \fig{amp} for notation):
 \beq \label{NEWSP}
 G_{3\pom}\Big{/} g_i(b = 0 )\,\ll\,\,1;~~~~~~~~ m\,\gg\, m_1 
~\mbox{and}~m_2
 \eeq
 
 The second equation in \eq{NEWSP} leads to the fact that $b''$ in 
\eq{EAPR} is much
 smaller than $b$ and $ b'$,
  therefore, \eq{EAPR} can be re-written in
 a simpler form
 \bea \label{EAPR1}
 \Omega_{i,k}(r_\bot, Y - Y_0, b)\,\,&=&\,\Bigg(\int d^2 b''\,
G^{\mbox{\tiny dressed}}\Lb
 T\Lb r_\bot, Y - Y_0, \vec{b}''\Rb\Rb\Bigg)\,\int d^2 b' g_i\Lb
 \vec{b}'\Rb \,g_k\Lb
 \vec{b} - \vec{b}'\Rb \,\nn\\
 &=&\,\tilde{G}^{\mbox{\tiny dressed}}\Lb r_\bot, Y - Y_0\Rb\,\,
\int d^2 b' g_i\Lb
 \vec{b}'\Rb \,g_k\Lb \vec{b} - \vec{b}'\Rb 
\eea

Using the first small parameter of \eq{NEWSP}, we  see 
 that the main contribution stems from the net diagrams shown in \fig{amp}-b.
 The sum of these diagrams\cite{GLM2CH} leads to the following expression 
for $
 \Omega_{i,k}(s,b)$
 \bea \label{OMEGA}
\Omega\Lb r,  Y-Y_0; b\Rb~~&=& ~~ \int d^2 b'\,
\,\,\,\frac{ g_i\Lb\vec {b}'\Rb\,g_k\Lb\vec{b} -
 \vec{b}'\Rb\,\tilde{G}^{\mbox{\tiny dressed}}\Lb r, Y - Y_0\Rb
}
{1\,+\,G_{3\pom}\,\tilde{G}^{\mbox{\tiny dressed}}\Lb r, Y - Y_0\Rb\left[
g_i\Lb\vec{b}'\Rb + g_k\Lb\vec{b} - \vec{b}'\Rb\right]} ;\label{OM}\\
g_i\Lb b \Rb~~&=&~~g_i \,S_p\Lb b; m_i \Rb ;\label{g}
\eea
where
\beq \label{SB}
S_p\Lb b,m_i\Rb\,=\,\frac{1}{4 \pi} m^3_i \,b \,K_1\Lb m_i b
 \Rb~~~\xrightarrow{\mbox{Fourier image} }~~~\frac{1}{\Lb 1
 + Q^2_T/m^2_i\Rb^2}
\eeq
\beq \label{GTILDE}
\tilde{G}^{\mbox{\tiny dressed}}\Lb r, Y -Y_0\Rb\,\,=\,\,\int d^2 b
 \,\,G^{\mbox{\tiny dressed}}\Lb T\Lb r, Y - Y_0, b\Rb\Rb
 \eeq
where $ T\Lb r, Y - Y_0, b\Rb$ is given by \eq{T}.

The impact parameter  dependence  of $S_p\Lb b,m_i\Rb$ is purely 
phenomenological,
 however, \eq{SB} which has a form of the electromagnetic proton  form 
factor,
 leads to the correct ($\exp\Lb - \mu b\Rb$) behavior at large 
$b$\cite{FROI},
 and has correct behavior at large $Q_T$, which has been calculate in the
 framework of perturbative QCD \cite{BRLE}. We wish to draw the reader's 
attention to the fact 
 that $m_1$ and $m_2$ are the two dimensional scales in a hadron, which in
 the framework of the constituent quark model,  we assign to the size of 
the
 hadron ($R_h \propto 1/m_1$), and the size of the constituent quark
 ($R_Q 
\propto 1/m_2$).

Note  that  $\tilde{G}^{\mbox{\tiny dressed}}\Lb Y - Y_0\Rb$ does not 
depend
 on $b$.  In all previous formulae, the value of the triple BFKL Pomeron
 vertex
 is known: $G_{3 \pom} = 1.29\,GeV^{-1}$.
  
\begin{table}[h]
\begin{tabular}{|l|l|l|l|l|l|l|l|l|}
\hline
model &$\lambda $ & $\phi_0$ ($GeV^{-2}$) &$g_1$ ($GeV^{-1}$)&$g_2$
 ($GeV^{-1}$)& $m(GeV)$ &$m_1(GeV)$& $m_2(GeV)$ & $\beta$ \\
\hline
I(soft int.)& 0.38& 0.0019 & 110.2&  11.2 & 5.25&0.92& 1.9 & 0.58  \\
\hline
II:(soft + DIS)& 0.38& 0.0022 & 96.9&  20.96 & 5.25&0.86& 1.76 & 0.66
  \\
 \hline 
  
  \end{tabular}
\caption{Fitted parameters of the model. Fit I: parameters for the
 soft interaction at high energy are  taken 
from Ref.\cite{GLM2CH}. The additional parameters for DIS were
 found by fitting to the $F_2$ structure function (see below).
 Fit II: joint fit to the soft interaction data  
 at high energy and 
the DIS data.}
\label{t1}
\end{table}

For  further discussion, we introduce the notation 

 \beq \label{NBK}
N^{BK}\Lb G^i_\pom\Lb r_\bot, Y,b \Rb\Rb \,\,=\,\,a\,\Lb 1
 - \exp\Lb -  G^i_\pom\Lb r_\bot, Y, b\Rb\Rb\Rb\,\,+\,\,\Lb 1 - a\Rb
\frac{ G^i_\pom\Lb  r_\bot, Y, b\Rb}{1\,+\, G^i_\pom\Lb r_\bot, Y, b\Rb},
\eeq 
 with $a = 0.65$ .
 \eq{NBK} is an analytical approximation to the numerical solution for  
the 
BK equation\cite{LEPP}. $G^{i}_\pom\Lb  r_\bot, Y; b\Rb \,=\,\,
 g_i\Lb b \Rb \,\tilde{G}^{\mbox{\tiny dressed}}\Lb r_\bot, Y - Y_0\Rb $.
 We recall that the BK equation sums the `fan'  diagrams.
 
  \subsection{ Results of the fit}
  
In this paper we make two fits. In the first one (fit I in Table 1 and Table
 3) we do not change the parameters that govern the soft interactions in 
our
 model, and  are shown in Table 1. The additional parameters that we 
need
 for  the description of the deep inelastic data, and which we will 
discuss in the
 next section (see Table 3),  were fitted using the HERA data on the deep
 inelastic structure function $F_2$. The second fit, is a joint fit to 
the
 soft strong interaction data and the DIS data.
 In \fig{dis} we show
the
 results of our model compared with the HERA data. The model predictions 
are in accord with the data for $0.85 \leq Q^{2} \leq 27 GeV^{2}$, while 
for 
higher values of $Q^{2}$ and of $x$, the model values
 are slightly larger than the data.

 In Table 2 we present 
our
 predictions for the soft interaction observables,  
in general the values obtained in the model for the soft interactions
agree with the published  LHC data, as well as the new preliminary TOTEM 
values at W =2.7,7,8,13 TeV (see Ref.\cite{Csorgo}).
We are in very good agreement with the data for 
$\sigma_{tot}$, $\sigma_{el}$ and $B_{el}$. Regarding $\sigma_{sd}$ and 
$\sigma_{dd}$, a problem exists when attempting to compare with the 
experimetal results.
This is due to the difficulties of measuring diffractive events at 
LHC energies, the different experiments have  different cuts on the 
values of the diffractive mass measured, making it problematic when 
attempting to compare the model predictions with the experimental results. 
           
In Table 2 we show the results of the two fits, the results  
are close to one 
another,
  the main difference shows up only at high energies. Indeed, in fit I 
the
 cross section for single diffraction is equal to 14.9 mb, while in fit II
 this value is smaller (13.1 mb).  The smaller value of the
 diffraction cross sections is closer to TOTEM and CMS data.

\begin{center}
\begin{table}
\begin{tabular}{|l|l|l|l| l l |l l|}
\hline
W &$\sigma_{tot} $& $\sigma_{el}$(mb) &$B_{el}$&~~~single& diffraction~~ &~~~~double& diffraction~~~ \\
(TeV)   &  (mb)  &        (mb)              &          $(GeV^{-2})$& $\sigma^{\rm smd}_{\rm sd}$ (mb)  &$\sigma^{\rm lmd}_{\rm sd}$ (mb)& $\sigma^{\rm smd}_{\rm dd}$ (mb)&$\sigma^{\rm lmd}_{\rm dd}$ (mb)\\
\hline
0.576 &62.3(60.7)& 12.9(13.1)&15.2(15.17) & 5.64(4.12)& 1.85(1.79)& 0.7(0.39)&0.46 (0.50)\\
\hline
0.9 & 69.2(68.07) &15(15.05)&16(15.95) &6.254.67)& 2.39(2.35)& 0.77(0.46)&0.67(0.745) \\
\hline
1.8&79.2(78.76)&18.2(19.1)&17.1(17.12)&7.1(5.44)&3.35(3.28) & 0.89(0.56)&1.17 (1.30) \\
\hline
2.74 &85.5(85.44)&20.2(21.4)&17.8(17.86)&7.6(5.91)&4.07(4.02) &0.97(0.63)&1.62(1.79)\\
\hline
7 &99.8(100.64)&25(26.7)&19.5(19.6)&8.7(6.96)& 6.2(6.17)&1.15(0.814)&3.27(3.67)\\
\hline
8 & 101.8(102.8)&25.7(27.4)&19.7(19.82)&8.82(7.1)&6.55(6.56) &1.17(0.841)&3.63(4.05)\\
\hline
13 & 109.3(111.07)&28.3(30.2)&20.6(20.74)&9.36(7.64)& 8.08(8.11) & 1.27(0.942)&5.11(5.74)\\
\hline
14 & 110.5(111.97)&28.7(30.6)&20.7(20.88)&9.44(7.71)& 8.34(8.42) & 1.27(0.96) &5.4(6.06)\\
\hline
57 & 131.7(134.0)&36.2(38.5) &23.1(23.0)&10.85(9.15)&15.02(15.01) & 1.56(1.26) &13.7(15.6)\\
\hline
\end{tabular}
\caption{ The values of cross sections versus
 energy. $\sigma^{\rm smd}_{\rm sd}$  and $\sigma^{\rm smd }_{\rm dd}$
 denote the cross sections for  diffraction dissociation
 in the small mass region, for single and double diffraction, which stem
 from the Good-Walker mechanism. While  $\sigma^{\rm lmd}_{\rm sd}$  and 
$\sigma^{\rm lmd}_{\rm dd}$
 denote high mass  diffraction, coming from the dressed 
Pomeron
 contributions.  The predictions of  fit II,  are shown in brackets.}

\label{t2}
\end{table}
\end{center}
 
  \section{Deep inelastic scattering}
  
  \subsection{Generalities}
  

   %
  In this section, we compare our amplitude with the experimental data on
  deep inelastic scattering (DIS). In the framework of our approach,
  the observables of DIS can be re-written using
  \beq \label{DIS1}
N_{T,L}\Lb Q, Y; b\Rb \,\,=\,\,\int \frac{d^2 r}{4\,\pi} \int^1_0 d z \,|\Psi^{\gamma^*}_{T,L}\Lb Q, r, z\Rb|^2 \,N\Lb r, Y; b\Rb
\eeq
where $Y \,=\,\ln\Lb 1/x_{Bj}\Rb$ and $x_{Bj}$ is the Bjorken $x$.
 $z$ is the fraction of energy carried by quark.
$Q$ is  the photon virtuality. $b$ denotes  the impact parameter  
 for the scattering of the colorless dipole of size $r$ with the
 proton. $N\Lb r, Y; b\Rb$ is the scattering amplitude of this dipole,
 which in our model can be written in the following form:

 \beq \label{DIS2}
 N\Lb r, Y; b\Rb\,\,=\,\, \alpha^2 \,N_1^{BK}\Lb g_1\, S\Lb b, m_1\Rb \tilde{G}_\pom\Lb r; Y \Rb\Rb\,\,+\,\,\beta^2 \,N_2^{BK}\Lb g_2\, S\Lb b, m_2\Rb \tilde{G}_\pom\Lb r; Y \Rb\Rb 
 \eeq
 
 In \eq{DIS1} $|\Psi^{\gamma^*}_{T,L}\Lb Q, r, z\Rb|^2$ is the probability
 to find a dipole of size $r$ in a photon with the virtuality $Q$,
 and with transverse or longitudinal polarization. The wave functions are
 known (see Ref.\cite{KOLEB} and reference therein) and they are equal to
  the following expressions:
  \begin{align}
  (\Psi^*\Psi)_{T}^{\gamma^*} &=
   \frac{2N_c}{\pi}\alpha_{\mathrm{em}}\sum_f e_f^2\left\{\left[z^2+(1-z)^2\right]\epsilon^2 K_1^2(\epsilon r) + m_f^2 K_0^2(\epsilon r)\right\},\label{DISWFT}   
  \\
  (\Psi^*\Psi)_{L}^{\gamma^*}&
  = \frac{8N_c}{\pi}\alpha_{\mathrm{em}} \sum_f e_f^2 Q^2 z^2(1-z)^2 K_0^2(\epsilon r),
\label{DISWFL}
\end{align}

where $\epsilon^2\,=\,m^2_f + z(1-z) Q^2$.

 Finally, the physical observables take the form:
 
 \beq \label{DIS3}
 \sigma_{T,L}\Lb Q,Y\Rb\,\,=\,\,2\int d^2 b \,\, N_{T,L}\Lb Q, Y; b\Rb
 \eeq
 
 \beq \label{DIS4}
 F_2\Lb Q, Y\Rb\,\,=\,\,\frac{Q^2}{ 4 \pi^2 \alpha_{\rm \mbox{e.m.}}}\Big\{ \sigma_T\,+\,\sigma_L\Big\}
 \eeq
   
  
  \subsection{Modification  to also include  DIS}
  
  First, we need to include the mild violation of the geometric
 scaling behavior of the scattering amplitude. We use the same
 procedure as  has been suggested in Res.
 \cite{IIM,SATMOD0,SATMOD1, SATMOD2,SATMOD3, SATMOD4, SATMOD5,
 SATMOD6,SATMOD7, SATMOD8, SATMOD9,SATMOD10,SATMOD11,SATMOD12,
 SATMOD13, SATMOD14,SATMOD15,SATMOD16, SATMOD17, CLP,CLP1}: 
we change $\bar{\gamma} $ in \eq{T}
  \beq \label{DIS5}
\bar{\gamma} =1 \,-\,\gamma_{cr}\,\,\to\,\,1 \,-\,\gamma_{cr}\,\,-\,\,\frac{1}{2\,\kappa\,\lambda\,Y} \,\ln\ \Lb r^2 \,Q^2_s\Lb b \Rb \Rb\,\,=\,\,0.63\,\,-\,\,\frac{1}{2\,\kappa\,\lambda\,Y} \,\ln\ \Lb r^2 \,Q^2_s\Lb b \Rb \Rb
\eeq  
where  $\kappa \,=\,\chi''\Lb \gamma_{cr}\Rb/\chi'\Lb \gamma_{cr}\Rb = 9.8$. $\chi\Lb \gamma\Rb$ is the BFKL kernel which has the foliowing form
\beq \label{DISCHI}
\chi\Lb \gamma\Rb\,=\,2 \psi(1) - \psi(\gamma) - \psi( 1 - \gamma)~~~~~~\mbox{while}~~~~~~\frac{\chi\Lb \gamma_{cr}\Rb}{1 - \gamma_{cr}} = \frac{d \chi\Lb \gamma\Rb}{ d \gamma}\Bigg{|}_{\gamma = \gamma_{cr}}
\eeq
where $\psi(z) = d \Gamma(z)/d z $ is the Euler $\psi$-function 
(see Ref.\cite{RY} formula {\bf 8.360}).

Since we take into account the contribution of the heavy $c$-quark we
 introduce a correction due to large mass of this quark:
\beq \label{DIS6}
x_{Bj}\,\,\to\,\,x_{Bj}\,\Lb \frac{1}{1+ \frac{4 m^2_c}{Q^2}}\Rb~~~~~~~~~~~~~\mbox{or}~~~~~~~~~~Y_c = Y - \ln\Lb1 + 4\,m^2_c/Q^2\Rb\eeq

In describing the saturation phenomena and fitting the 
strong interaction data,
 we assumed that the QCD coupling is  frozen at some  value 
of momentum $\mu_{\rm soft}$. 
 However, for DIS we take into account the running QCD coupling,
 replacing \eq{DIS4} by the following expression

\beq \label{DIS7}
  F_2\Lb Q,Y\Rb\,\,=\,\,\frac{Q^2}{ 4 \pi^2 \alpha_{\rm \mbox{e.m.}}}\Bigg\{ \frac{\bas\Lb Q^2\Rb}{\bas\Lb \mu^2\Rb} \, \sigma^{\rm light \,q}\Lb Q, Y\Rb\,\,+\,\,\frac{\bas\Lb Q^2 + 4m^2_c \Rb}{\bas\Lb \mu^2\Rb}   \sigma^{\rm charm \,q}\Lb Q, Y_c\Rb\,\Bigg\}
  \eeq
  
  where $\mu$ denotes the typical mass of the soft strong interaction $\mu 
\sim 1\,GeV$ and 
  \beq \label{R}
  \frac{\bas\Lb Q^2\Rb}{\bas\Lb \mu^2\Rb}  \,\,=\,\,\frac{1}{1\,+\,\beta \bas\Lb \mu^2\Rb\,\ln\Lb Q^2/\mu^2\Rb}
  \eeq  
    with $\beta=3/4$.

    We consider the  strong interaction data for  energies 
$W \geq 
0.546\,TeV$, while the experimental data from HERA were measured for 
lower energies. Therefore, we need to include the
 contribution of the secondary Reggeons which give a substantial
 contribution\cite{SR}.

  \beq \label{REGGE}
  \sigma_{\reg}\Lb Q, Y\Rb  \,\,=\,\, \int \frac{d^2 r}{4\,\pi} \Bigg\{  (\Psi^*\Psi)_{T}^{\gamma^*} \,+\,  (\Psi^*\Psi)_{L}^{\gamma^*} \Bigg\} \,A_\reg\,r^2 \,\Lb \frac{Q^2}{x_{Bj}\,Q^2_0} \Rb^{ \alpha_\reg\Lb 0 \Rb    -  1}
  \eeq
  with
  $Q_0 = 1  \,GeV$.
  
  The final equation for $F_2$ takes a form:
  \beq \label{DIS8}  
  F_2\Lb Q,Y\Rb\,\,=\,\,\frac{Q^2}{ 4 \pi^2 \alpha_{\rm \mbox{e.m.}}}\Bigg\{ \frac{\bas\Lb Q^2\Rb}{\bas\Lb \mu^2\Rb} \, \sigma^{\rm light \,q}\Lb Q, Y\Rb\,\,+\,\,\frac{\bas\Lb Q^2 + 4m^2_c \Rb}{\bas\Lb \mu^2\Rb}   \sigma^{\rm charm \,q}\Lb Q, Y_c\Rb\,\,+\,\, \sigma_{\reg}\Lb Q, Y\Rb\Bigg\}
  \eeq  
  
  
  \subsection{The description of the HERA data}
  
   We introduce  in \eq{DIS8},  a set of new parameters  for DIS: 
$m_q$-mass 
of the light quark, which we hope will be of the order of the constituent
 quark mass ($\sim 300\,MeV$), the mass of charm quark ($m_c = 1.2 \div 1.5 
\,GeV$), $\mu$ which we believe will be of the order of 1\,GeV, and
  we introduce two new parameters $A$ and $\alpha_\reg\Lb 0 \Rb$ for the
 secondary Reggeon contribution. For $A$ there is only one restriction 
that at $x_{bj} = 4 \,10^{-6}$  $ \sigma^{\rm light \,q}_{\reg}  \leq 0.02
 \sigma_{tot}$, while $ \alpha_\reg\Lb 0 \Rb = 0.4 \div 0.6$.
        
\begin{table}[h]
\begin{center}
\begin{tabular}{||l|l|l|l|l|l|l||}
\hline
model &$m_q (GeV$ & $m_c(GeV)$  & $\as(\mu)$&$\mu$
 ($GeV$)&  $A_\reg(GeV^2$ &$\alpha_\reg(0))$\\
\hline
I & 0.3 & 1.25 &0.263& 1.2 & 2.34 & 0.55  \\
\hline
 II & 0.2 & 1.2 &0.34& 1.25 & 5.44 & 0.56  \\\hline
\end{tabular}
\end{center}
\caption{Fitted parameters for DIS. The description of fit I and fit II is
 given in section 2.3, and in the caption of Table 1.}
\label{t2}
\end{table}

\begin{figure}[ht]
    \centering
  \leavevmode
      \includegraphics[width=14cm]{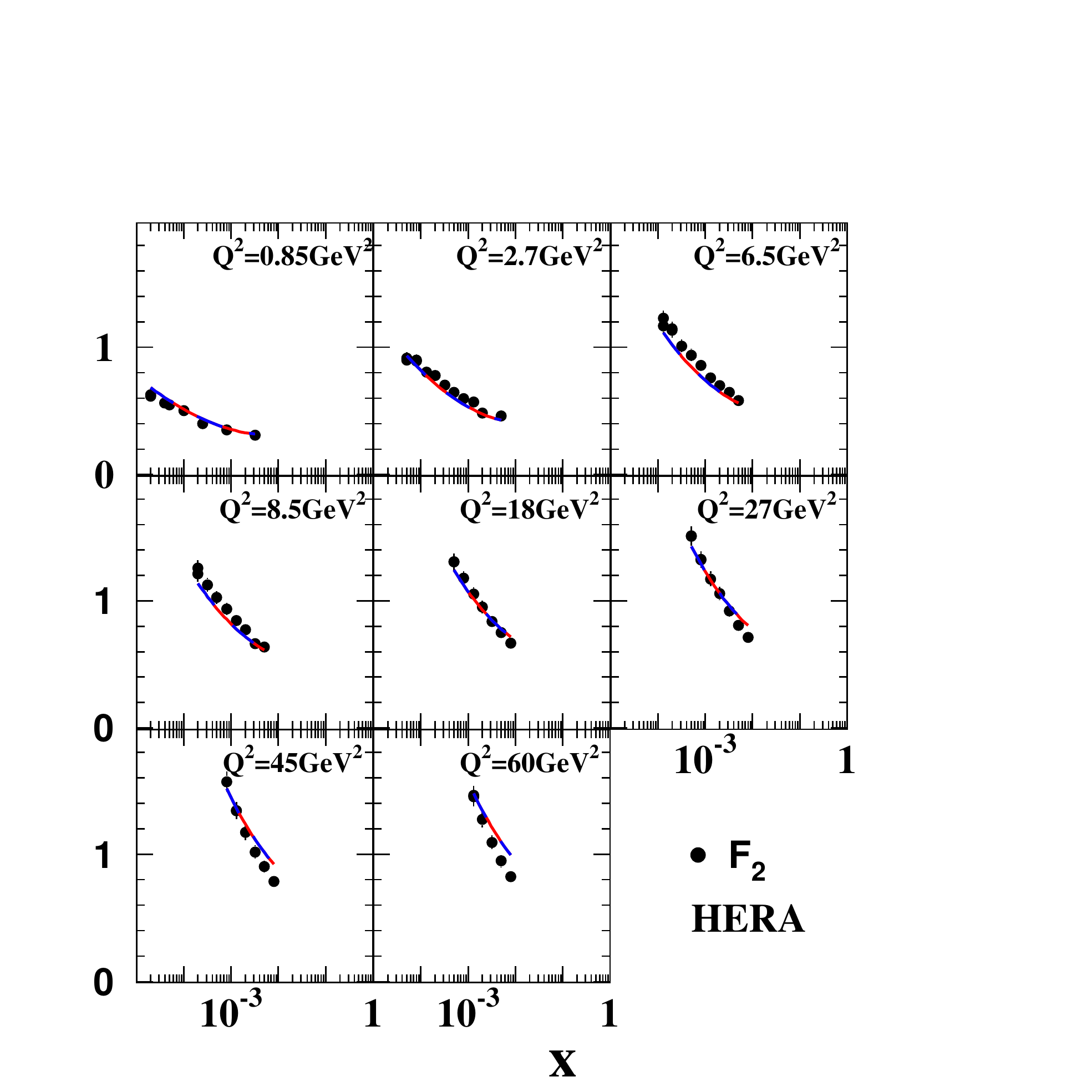}  
      \caption{$F_2$ versus $x$ at fixed $Q$. The red curve corresponds to
 fit I, while the blue one describes  fit II. Data is taken from
 Ref.\cite{HERA}.}
\label{dis}
   \end{figure}

In Table 3 we  display the parameters that were  determined by 
fitting to the data,  \fig{dis} 
shows the quality of our  fit to  the DIS  HERA data.

 We consider the fit shown in \fig{dis} to be in very good agreement 
with the experimental data,
  and to demonstrate that our model is able to describe the hard
 processes to within an accuracy of 5\%.
 
  
  \section{Inclusive production}

 The cross section of the inclusive production is  a very important
 observable for our estimates, since it  indicates   how well, we can 
describe
 the multi particle generation processes in our model.  We have described
 the experimental data in our soft interaction model \cite{GLMINCL},
 we now  recalculate using our generalization of the
 model, that we have discussed above.   Ref.\cite{SATMOD7} showed
 that the CGC/saturation approach is able to describe the LHC data on
 inclusive production. In this section we re-visit these calculations,
 using our model, which we can now apply both to soft and to hard 
processes.
       
\begin{figure}[ht]
    \centering
  \leavevmode
      \includegraphics[width=14cm]{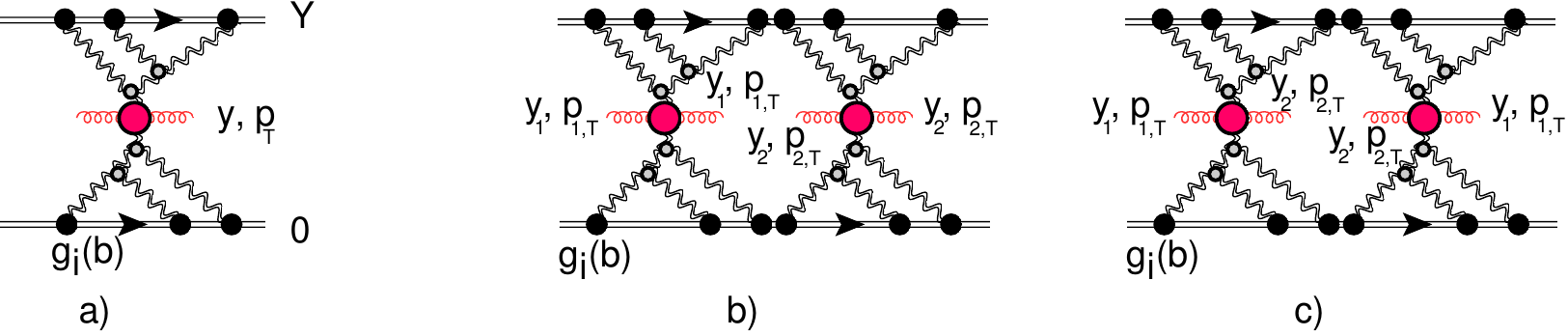}  
      \caption{The generic Mueller  diagrams\protect\cite{MUDIA} for single
  inclusive(\protect\fig{incl}-a) and for double inclusive
 (\protect\fig{incl}-b and 
      \protect\fig{incl}-c) production. \protect\fig{incl}-b describes
 the double inclusive cross section, while \protect\fig{incl}-c shows
 the interference diagram for the Bose-Einstein correlation. For ease
 of drawing we take $y_1=y_2$.}
\label{incl}
   \end{figure}

 
 The expression for the inclusive cross section takes the
 form\cite{KTINC,SATMOD7,KOLEB} (see \fig{incl} for notation)

 \bea \label{INC1}
\frac{d \sigma}{d y \,d^2 p_{T}}\,\,& & \frac{2C_F}{\alpha_s (2\pi)^4}\,\frac{1}{p^2_T}\int d^2 \vec b \,d^2 \vec B \,d^2 r \,e^{i \vec{p}_T\cdot \vec{r}}\,\,\nabla^2_T\,N^{h_1}_G\Lb Y - y; r; b \Rb\,\,\nabla^2_T\,N^{h_2}_G\Lb y; r; |\vec b-\vec B| \Rb. 
\eea 
where the  scattering amplitudes $N^{h_i}_G$ can be found from the dipole amplitude\cite{KTINC}
 \beq \label{INC2}
N^{h_i}_G\Lb y_i ; r; b \Rb\,\,=\,\,2 \,N\Lb y_i ; r; b \Rb\,\,-\,\,N^2\Lb y_i ; r; b \Rb,
\eeq 
and $r$ denotes the dipole size. $C_F = (N^2_c-1)/2N_c$. For further
 discussion it is convenient to introduce
two more observables
\bea \label{INC3}
\frac{d \sigma_{i,j}}{d y \,d^2 p_{T}\,d^2B }\,\,&=& \frac{2C_F}{\alpha_s (2\pi)^4}\,\frac{1}{p^2_T}\int d^2  b \, \,d^2 r \,e^{i \vec{p}_T\cdot \vec{r}}\,\,\nabla^2_T\,N^{i}_G\Lb Y - y; r; b \Rb\,\,\nabla^2_T\,N^{j}_G\Lb y; r; |\vec b-\vec B| \Rb;\nn\\
\frac{d \sigma_{i,j}}{d y \,d^2 p_{T}\,d^2 B\,d^2 b }\,\,&= &\frac{2C_F}{\alpha_s (2\pi)^4}\,\frac{1}{p^2_T}\int  \,d^2 r \,e^{i \vec{p}_T\cdot \vec{r}}\,\,\nabla^2_T\,N^{i}_G\Lb Y - y; r; b \Rb\,\,\nabla^2_T\,N^{j}_G\Lb y; r; |\vec b-\vec B| \Rb
\eea 
where 
\beq \label{INC31}
N^{i}_G\Lb y_i ; r; b \Rb\,\,=\,\,2 \,N_i^{BK}\Lb g_i\, S\Lb b, m_i\Rb \tilde{G}_\pom\Lb r; y_i \Rb\Rb\,\,-\,\,\Lb N_i^{BK}\Lb g_i\, S\Lb b, m_i\Rb \tilde{G}_\pom\Lb r; y_i \Rb\Rb\Rb^2
\eeq

\begin{figure}[ht]
    \centering
  \leavevmode
      \includegraphics[width=10cm]{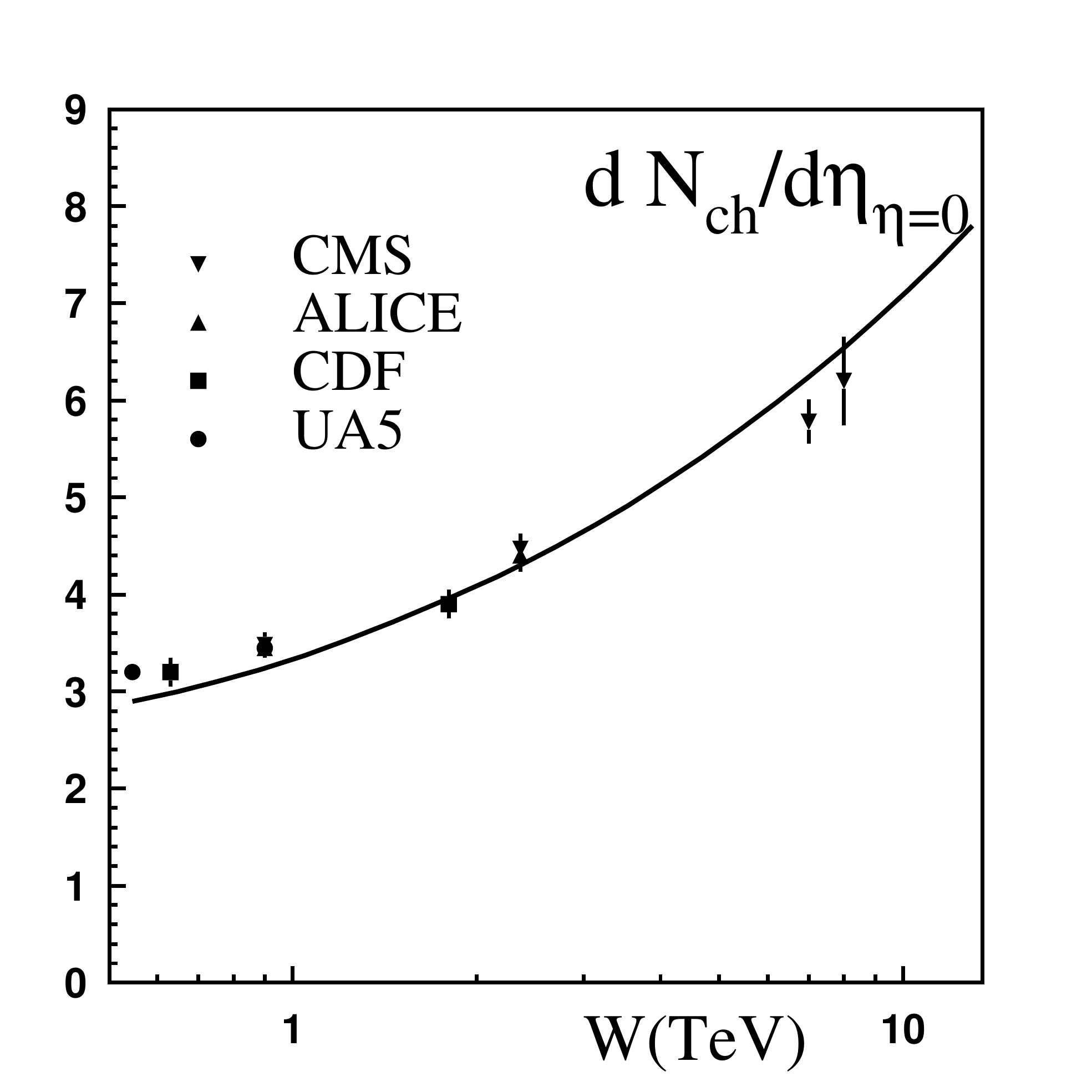}  
      \caption{$\frac{d N}{d y}\Big{|}_{y = 0}$ versus energy W.
 The experimental data are taken from Refs.\cite{ALICEIN,CMSIN,ATLASIN}
 and Ref.\cite{PDG}}
\label{dndy}
   \end{figure}

Taking for $N\Lb y_i ; r; b \Rb$ in \eq{INC2}  the amplitude of \eq{DIS2}
 we  obtain
\beq 
\label{INC4}
\frac{d N}{d y}\Big{|}_{y = 0}\,\,=\,\,\frac{1}{\sigma_{\rm NSD}}\int d^2
 p_T \frac{d \sigma}{d y \,d^2 p_{T}}
\eeq
The values of  $\sigma_{\rm NSD} \,=\,\sigma_{\rm tot} - \sigma_{\rm el} 
- \sigma_{\rm single\,diffraction}$ we take from the description of total
 and diffraction cross section in our model\cite{GLM2CH}.  One can see that
 integral over $p_T$ is logarithmical divergent  at small $p_T$. As 
 shown in Ref.\cite{KLN}  this divergence is regularized by the mass
 of produced gluon jet at $y=0$. In \fig{dndy} we plotted our estimates
 for $\frac{d N}{d y}\Big{|}_{y = 0}$ using the value of this mass as was
 taken in Ref.\cite{SATMOD7} $m_{\rm jet} = 350 \,MeV$. The agreement with
 the experimental data is good and it gives us confidence that our model is
 able to discuss the typical process of many particle production. We do not
 need to discuss the rapidity distribution of the single inclusive cross
 section, since it has been discussed in Refs.\cite{GLMINCL,SATMOD7},
 where it is shown that this distribution agrees with the experimental 
data.

  \section{Azimuthal angle correlations}
  
  \subsection{Double inclusive cross section}
 
 The Mueller diagram\cite{MUDIA} for double inclusive cross section
 is shown in \fig{incl}-b. Using \eq{INC3} this cross section can be
 written in the form:
 \bea
\Sigma_{i,j}\,\equiv\, \frac{d^2 \sigma_{i,j}}{d y_1 \,d^2 p_{1,T}\,d y_2 \,d^2 p_{2,T}} \,\,&=&\,\,\int d^2 B \frac{d \sigma}{d y_1 \,d^2 p_{1,T}\,d^2B } \,\frac{d \sigma}{d y_2 \,d^2 p_{2,T}\,d^2B } \label{DINC1}\\
  \frac{d^2 \sigma}{d y_1 \,d^2 p_{1,T}\,d y_2 \,d^2 p_{2,T}} \,\,&=&\alpha^4\, \Sigma_{1,1}  \,\,+\,\,2\,\alpha^2\,\beta^2\,  \Sigma_{1,2}  \,\,+\,\,\beta^4\,\Sigma_{2,2}\label{DINC11} 
    \eea
 We can re-write \eq{DINC1} in a different form if we introduce 
 \beq \label{DINC2}
 I^G_{i}\Lb y, r, Q_T\Rb\,\,=\,\,\int d^2 b \,e^{i \vec{Q}_T \cdot \vec{r}}\,\nabla^2_T\,N^{i}_G\Lb Y - y; r; b \Rb
 \eeq 
 Note, that $\vec{Q}_T$ denotes the transverse momentum carried by the
 BFKL Pomeron, which emits  gluons with momentum $\vec{p}_{1,T}$ or
 $\vec{p}_{2,T}$.
 
Plugging  \eq{DINC2} in \eq{DINC1} the expression for the double inclusive
 cross section takes the form
\bea \label{DINC3}
 \frac{d^2 \sigma_{i,j}}{d y_1 \,d^2 p_{1,T}\,d y_2 \,d^2 p_{2,T}} \,\,&=&\,\,\frac{2C_F}{\alpha_s (2\pi)^4}\,\frac{1}{p^2_{1,T}}\, 
 \frac{2C_F}{\alpha_s (2\pi)^4}\,\frac{1}{p^2_{2,T}}\int d^2 r_1\,  e^{ \vec{p}_{1,T} \cdot \vec{r}_1}\,\int d^2 r_2 \, e^{ \vec{p}_{2,T} \cdot \vec{r}_2}\\
  &\times&\,\,\int  \frac{d^2 Q_T}{(2 \pi )^2}  I^G_{i}\Lb Y - y_1, r_1, Q_T\Rb\, I^G_{i}\Lb Y - y_2, r_2, Q_T\Rb  I^G_{j}\Lb  y_1, r_1, Q_T\Rb\, I^G_{j}\Lb  y_2, r_2, Q_T\Rb \nn
 \eea
 Therefore, using either \eq{DINC1} or \eq{DINC3}  and the decomposition 
of
 \eq{DINC11},  one can calculate the double inclusive cross section.
 

  \subsection{Bose-Einstein correlation: energy dependence}
 The double inclusive cross section of two identical gluons has the
 following general form:
\beq \label{BEE1}
 \frac{d^2 \sigma}{d y_1 \,d y_2  d^2 p_{T1} d^2 p_{T2}}\Lb \rm identical\,\,
 gluons\Rb\,\,=\,\,
  \, \frac{d^2 \sigma}{d y_1 \,d y_2  d^2 p_{T1} d^2 p_{T2}}\Lb \rm different
 \,\, gluons\Rb \Big( 1 \,+\,C\Lb L_c |\vec{p}_{T2} -   \vec{p}_{T1}|\Rb\Big)
  \eeq
where $C\Lb L_c |\vec{p}_{T2} -   \vec{p}_{T1}|\Rb$  denotes the 
correlation
 function, and $L_c$   
 the correlation length. The first term in \eq{BEE1} is given by \eq{DINC1}
 or  \eq{DINC3}, while the second term describes the interference diagram
 for the identical gluons (see \fig{incl}-c and Refs. \cite{GOLE,GOLELAST}
 for details). The expression for the interference term is more 
transparent
 in momentum representation, where it  has the form
 \bea \label{BEE2}
 && \frac{d^2 \sigma_{i,j}\Lb \rm interference\,\,
contribution\Rb}{d y_1 \,d y_2  d^2 p_{T1} d^2 p_{T2}}\,=\,\,\frac{1}{N^2_c - 1}\,\frac{2C_F}{\alpha_s (2\pi)^4}\,\frac{1}{p^2_{1,T}}\, 
 \frac{2C_F}{\alpha_s (2\pi)^4}\,\frac{1}{p^2_{2,T}}\int d^2 r_1\,  e^{ \vec{p}_{1,T} \cdot \vec{r}_1}\,\int d^2 r_2 \, e^{ \vec{p}_{2,T} \cdot \vec{r}_2}\nn\\
 &&  \times \Bigg\{
 \,\int  \frac{d^2 Q_T}{(2 \pi )^2}  I^G_{i}\Lb Y - y_1, r_1, Q_T\Rb\, I^G_{i}\Lb Y - y_2, r_2, Q_T\Rb  I^G_{j}\Lb  y_1, r_1, \vec{Q}_T - \vec{p}_{12,T} \Rb\, I^G_{j}\Lb  y_2, r_2,\vec{ Q}_T - \vec{p}_{12,T}\Rb \nn\\
 &&=\,\,\int  \frac{d^2 Q'_T}{(2 \pi )^2}  I^G_{i}\Lb Y - y_1, r_1, \vec{Q'}_T + \h \vec{p}_{12,T}\Rb\, I^G_{i}\Lb Y - y_2, r_2, \vec{Q'}_T + \h\vec{p}_{12,T}\Rb \nn\\
&& \times \,\,\, I^G_{j}\Lb  y_1, r_1, \vec{Q}_T -\h\vec{p}_{12,T} \Rb\, I^G_{j}\Lb  y_2, r_2,\vec{ Q}_T -\h\vec{p}_{12,T}\Rb\Bigg\}
 \eea
 \eq{BEE2} takes into account that the lower  BFKL Pomerons in
 \fig{incl}-c, carry momenta $\vec{Q}_T - \vec{p}_{12,T}$, where $
 \vec{p}_{12,T} \,\equiv\,\vec{p}_{1,T} \,-\,\vec{p}_{2,T}$.
 
 \eq{BEE2} can be re-written in the impact parameter representation using
 \eq{INC3}
 \bea \label{BEE3}
&& \frac{d^2 \sigma_{i,j}}{d y_1 \,d y_2  d^2 p_{T1} d^2 p_{T2}}\Lb \rm interference\,\,
contribution\Rb\,=\\
&&\frac{1}{N^2_c - 1}\int d^2  \tilde{b} \,e^{i \vec{p}_{12,T} \cdot \vec{\tilde{b}}}\int d^2 B' d^2 b \, \frac{d \sigma_{i,j}}{d y \,d^2 p_{T}\,d^2 B\,d^2 b } \Lb \vec{b} + \h \vec{\tilde{b}}, \vec{B'} + \h \vec{\tilde{b}} \Rb \,\frac{d \sigma_{i,j}}{d y \,d^2 p_{T}\,d^2 B\,d^2 b } \Lb \vec{b} - \h \vec{\tilde{b}}, \vec{B'} - \h \vec{\tilde{b}} \Rb  \nn
\eea 

\begin{figure}[ht]
    \centering
  \leavevmode
  \begin{tabular}{c c c}
      \includegraphics[width=7cm]{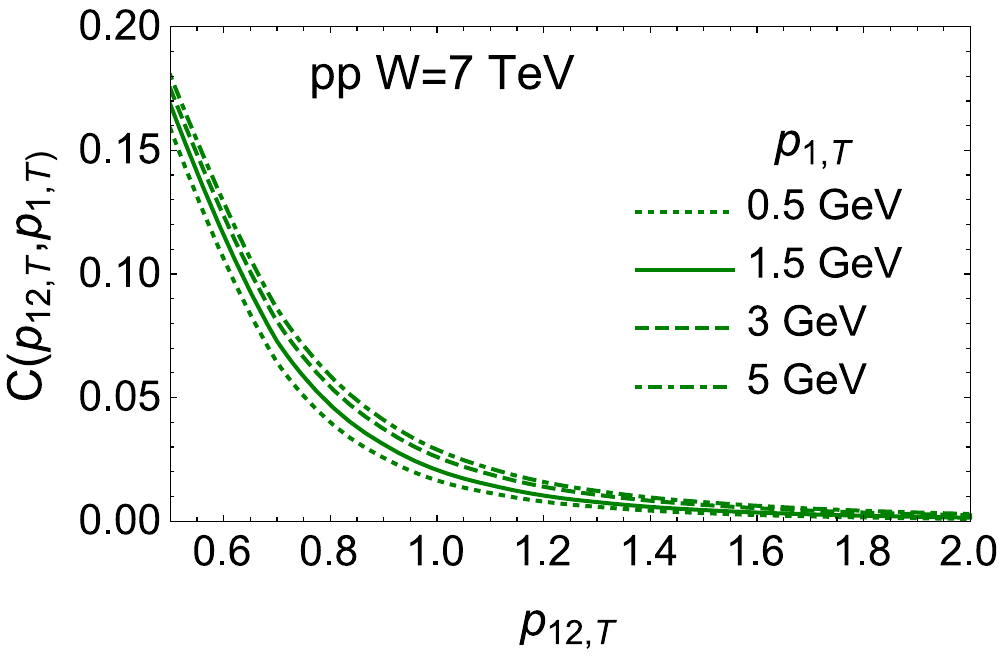}  &~~~~~~~~~~~&\includegraphics[width=7cm]{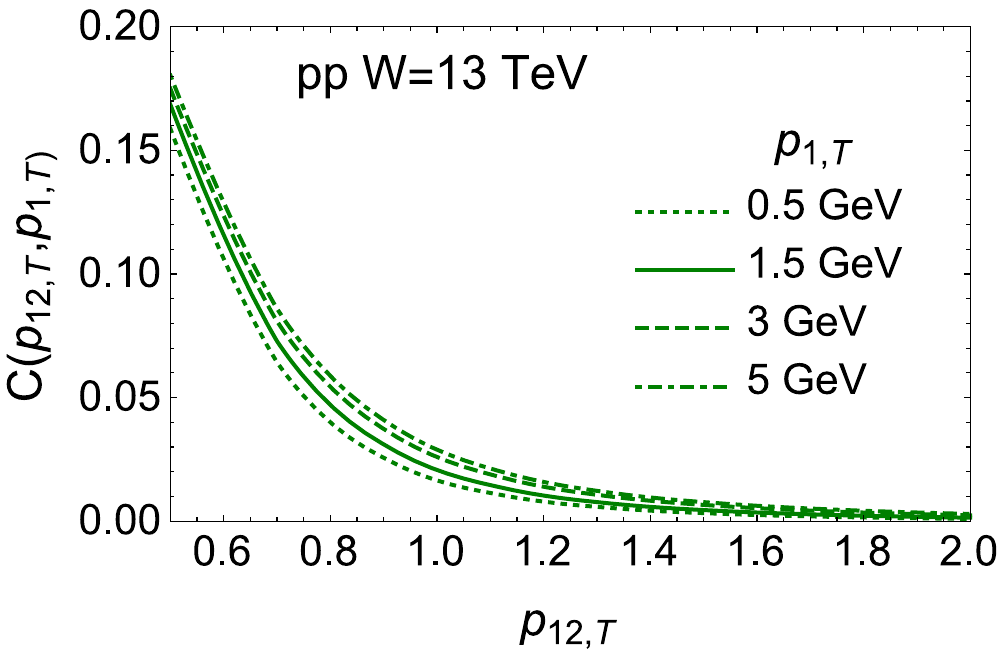} \\
      \fig{CW}-a&~~~~~~~~~~~& \fig{CW}-a\\
      \end{tabular}
      
 \caption{Correlation function $C\Lb L_c |\vec{p}_{T2} -   \vec{p}_{T1}|\Rb$ 
 versus $ \vec{p}_{12,T} \,\equiv\,\vec{p}_{1,T} \,-\,\vec{p}_{2,T}$  at
 different values of $p_{1,T}$ and  energies $W = 7 \,TeV$
 (\protect\fig{CW}-a) and  $W=13\,TeV$ ( 
 \protect\fig{CW}-a).}
 \label{CW}
   \end{figure}

Finally, using the decomposition of \eq{DINC11}, we can calculate the
 correlation function.
In \fig{CW} we show the calculated correlation function
\beq \label{C}
C\Lb L_c |\vec{p}_{T2} -   \vec{p}_{T1}|\Rb\,\,=\,\,\frac{\frac{d^2 \sigma}{d y_1 \,d y_2  d^2 p_{T1} d^2 p_{T2}}\Lb \rm interference\,\,
contribution\Rb}{ \frac{d^2 \sigma}{d y_1 \,d y_2  d^2 p_{T1} d^2 p_{T2}}\Lb \rm different
 \,\, gluons\Rb}
 \eeq

 From this figure we  note that the correlation function does not
 depend on energy. This is an expected result. Indeed, the production
 of two parton showers, which is taken into account in \fig{incl}-b and
 \fig{incl}-c, leads to the correlation function, which does not depend
 on $y_{12} = |y_1 - y_2|$ (long range rapidity correlations(LRC)). 
 This happens in our approach where the structure of one parton shower
 cannot be reduced to the exchange of the one BFKL Pomeron. \fig{CW}
 illustrates that the dependence on energy also cancels in the ratio of \eq{C}.

 \begin{boldmath}
  
  \subsection{Bose-Einstein correlation:  values of $v_n$ and its
 multiplicity dependence }
   \end{boldmath}  
 We first  introduce $v_n$,  that can be defined it terms of 
the following representation of the double inclusive cross section
  \beq \label{BEN1}
     \frac{d^2 \sigma}{d y_1 \,d y_2  d^2 p_{T1} d^2 p_{T2}}\,\,\propto\,\,1
 \,\,+\,\,2 \sum_n v_{ n,n } \Lb p_{T1}, p_{T2}\Rb \,\cos\Lb n\,
\varphi\Rb
     \eeq
     where $ \varphi$ is the angle between    $\vec{p}_{T1}$ and
 $ \vec{p}_{T2} $.
     $v_n$ is determined  from  $v_{n,n}   \Lb p_{T1}, p_{T2}\Rb $
     \beq \label{vn}  
 1.~~    v_n\Lb p_T\Rb\,\,=\,\,\sqrt{v_{n,n}\Lb p_T, p_T\Rb}\,;\,
~~~~~~~~~~~~~2.~~~~  v_n\Lb p_T\Rb\,\,=\,\,\frac{v_{n, n}\Lb p_T,
 p^{\rm Ref}_T\Rb}{\sqrt{v_{n,n}\Lb p^{\rm Ref}_T, p^{\rm Ref}_T\Rb}}\,;
     \eeq
 \eq{vn}-1 and \eq{vn}-2  depict   two methods  of how the  values
 of $v_n$ have been extracted from the experimentally measured
  $v_{n,n} \Lb p_{T1}, p_{T2}\Rb$. Where $ p^{\rm Ref}_T$ denotes the
 momentum of the reference trigger.
     These two definitions are equivalent if  $v_{n, n}\Lb p_{T1},
 p_{T2}\Rb $ can be factorized as $v_{n, n}\Lb p_{T1}, p_{T2}\Rb\,=\,
     v_n\Lb p_{T1}\Rb\,v_n\Lb p_{T2}\Rb $. In this paper we use
 \eq{vn}-1 definition.
     
\begin{figure}[ht]
    \centering
  \leavevmode
      \includegraphics[width=10cm]{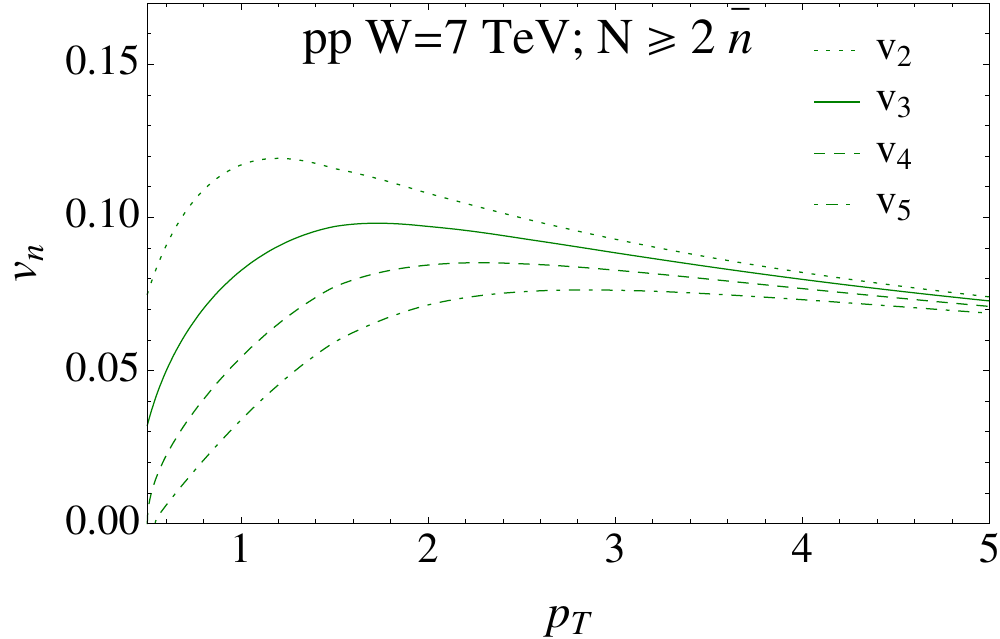}  
      \caption{$v_n$ versus $p_T$ for the proton-proton scattering at $W = 7\,TeV$.}
\label{vni}
   \end{figure}

     Taking into account \eq{C} and \eq{BEN1} we obtain
     \beq \label{BEN2}
     v_{n,n} \,\,=\,\,\frac{\int^{2 \pi}_0  d \varphi \,C\Big( 2 \,p_T\, \sin\Lb \h\varphi\Rb\Big)\,\cos\Lb n\,\varphi\Rb}{2\,\pi\,C\Lb p_{12,T}=0\Rb + \int^{2 \pi}_0 
      d \varphi \,C\Big( 2 p_T \sin\Lb \h\varphi\Rb\Big)\,\cos\Lb n\,\varphi\Rb}; ~~~~~~v_n\,=\,\sqrt{v_{n,n}\,};
      \eeq
 \eq{BEN2} gives the prescription for the calculation of  $v_n$ that is 
measured  as a sum of the events with all possible multiplicities of
 the secondary hadrons. However,  in practice,  only events with 
multiplicities
 larger than $2 \bar{n}$, where $\bar{n} $ is the average multiplicity 
which are
  measured in  single inclusive experiments. \fig{vni} shows our
 calculations for W= 13\,TeV.

     
 The dependence of $v_n$ on the multiplicity of the event has been
 discussed in Ref.\cite{GOLELAST}. \eq{BEE3} takes a different form:
 \bea \label{BEN3}
&& \frac{d^2 \sigma}{d y_1 \,d y_2  d^2 p_{T1} d^2 p_{T2}}\Lb \rm
 interference\,\,
contribution\Rb\,=\,\,\frac{1}{N^2_c - 1}\int d^2  \tilde{b} \,e^{i \vec{p}_{12,T} \cdot \vec{\tilde{b}}}\\
&& \times \,\,\,\int d^2 B' d^2 b \, \frac{d \sigma}{d y \,d^2 p_{T}\,d^2 B\,d^2 b } \Lb \vec{b} + \h \vec{\tilde{b}}, \vec{B'} + \h \vec{\tilde{b}} \Rb \,\frac{d \sigma}{d y \,d^2 p_{T}\,d^2 B\,d^2 b } \Lb \vec{b} - \h \vec{\tilde{b}}, \vec{B'} - \h \vec{\tilde{b}} \Rb \frac{\sigma^{(m)}\Lb \vec{b} + \vec{B'}\Rb}{\sigma_0}; \nn\\
&& \mbox{with} ~~~~~~~~\frac{\sigma^{(m)}\Lb \vec{b} + \vec{B'}\Rb}{\sigma_0}\,\,=\,\,\frac{\Gamma\Lb m - 2, 2 \Omega\Lb Y, \vec{b} + \vec{B'}\Rb\Rb}{\Gamma\Lb m - 2\Rb}; \label{BEN31}
\eea 
 where $\Omega\Lb r=1/m Y, b\Rb$  is given by \eq{OMEGA}. \eq{BEN3}
 describes the Bose-Einstein correlations in the event whose multiplicity
 is large than $ 5 \bar{n}$ ($N \geq 5 \bar{n}$), where $\bar{n}$ is the
 average multiplicity. 
  
     
       \begin{figure}[ht]
    \centering
  \leavevmode
      \includegraphics[width=10cm]{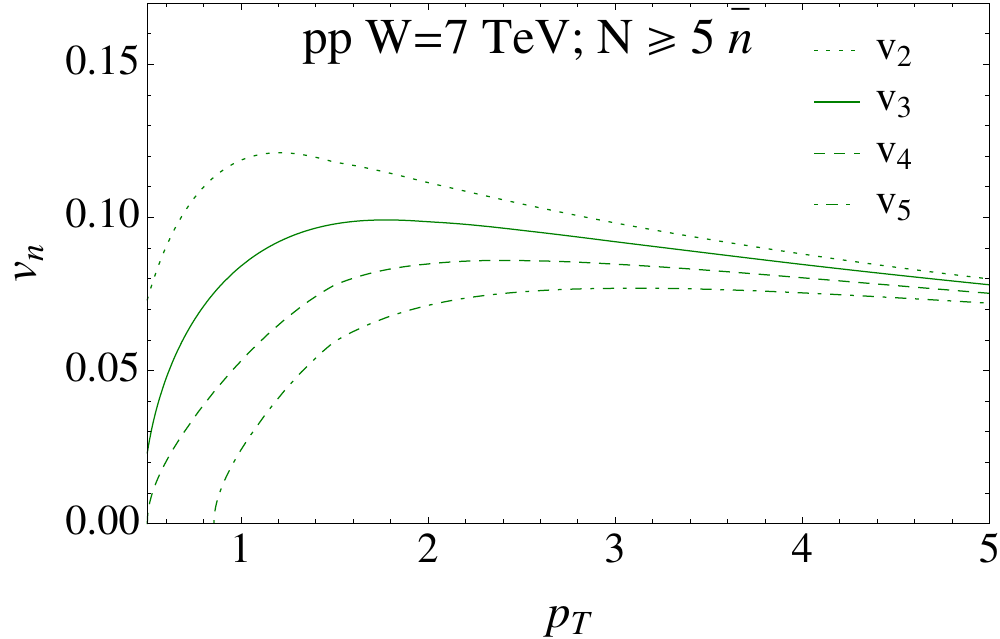}  
      \caption{$v_n$ versus $p_T$ for proton-proton scattering
 at $W = 7\,TeV$ for the multiplicities $N \geq\,5 \bar{n}$.}
\label{vn5}
   \end{figure}

\begin{figure}[ht]
    \centering
  \leavevmode
      \includegraphics[width=10cm]{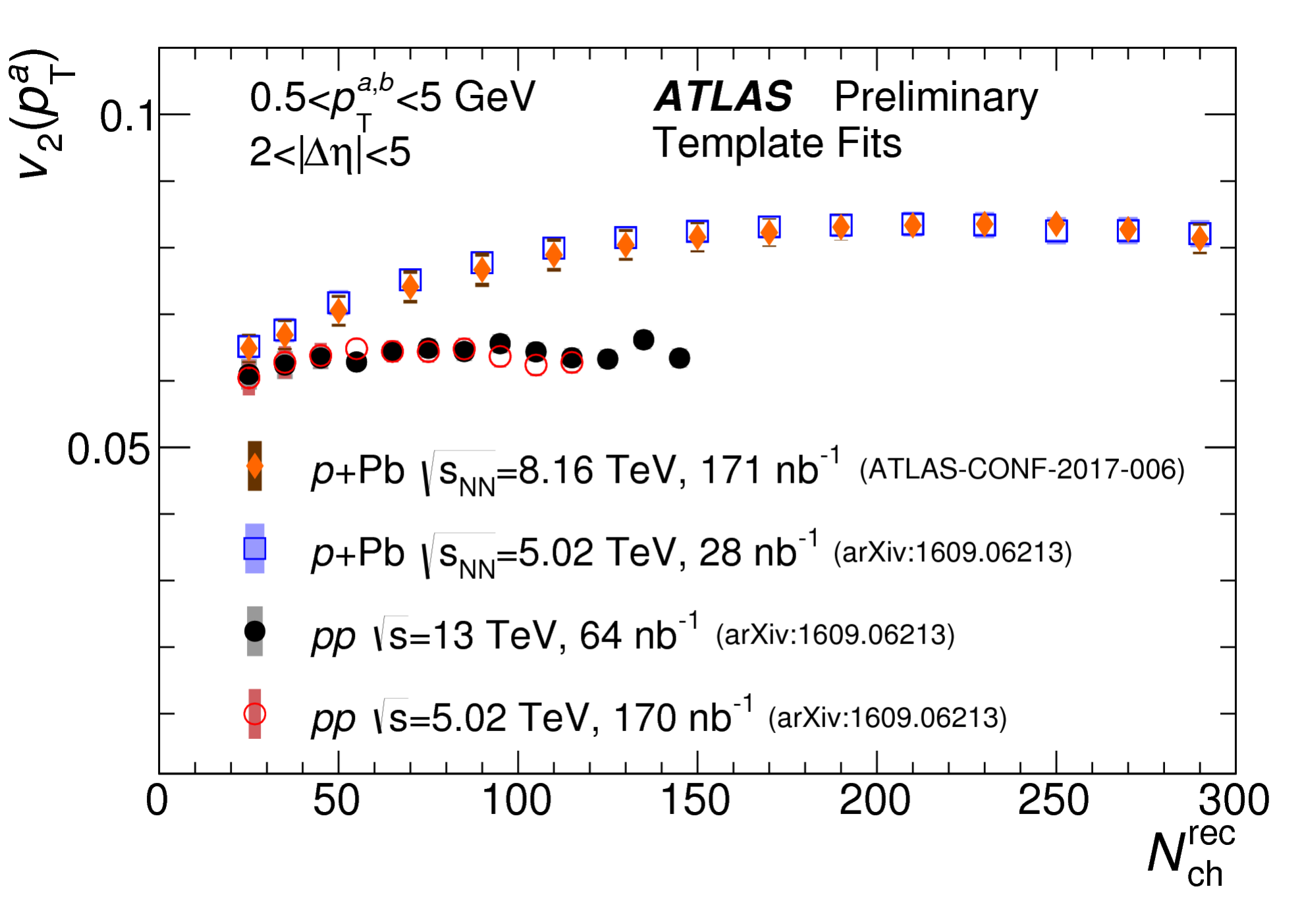}  
      \caption{$v_n$ versus multiplicities  for
 hadron-hadron and hadron-nucleus interactions.}
\label{exp}
   \end{figure}

 
Comparing \fig{vni} and \fig{vn5} one can see that in the framework of our 
approach,
 $v_n$ do not depend
 on the multiplicity of the event.
  This independence is in excellent agreement with the experimental 
data
 (see   Ref.\cite{ATLASPP} and \fig{exp}).
  Note, that $v_n$ do not depend on $N$ only for proton-proton scattering,
 while for hadron-nucleus collisions, such dependence is
  considerable.
  
     
 \fig{CW} shows  that  the correlation length $L_c \approx 1/m_1$
 (the typical momentum is about $m_1$). From Table 1,  the technical 
reason for
 this is clear, since the component with such characteristic
 momentum  makes the largest contribution. In more general language 
the
 correlation length 
  depends on the non-perturbative hadron structure. In terms of the 
processes,
 this typical transverse momentum is responsible for  diffractive scattering
  with the production of hadrons with small masses. Intuitively, we 
expect that
 diffractive production of large masses, which depend on the saturation 
scale,
 can lead to larger typical momenta (smaller correlation length). We will
 discuss these processes in the next section.
  
\begin{figure}[ht]
    \centering
  \leavevmode
      \includegraphics[width=12cm]{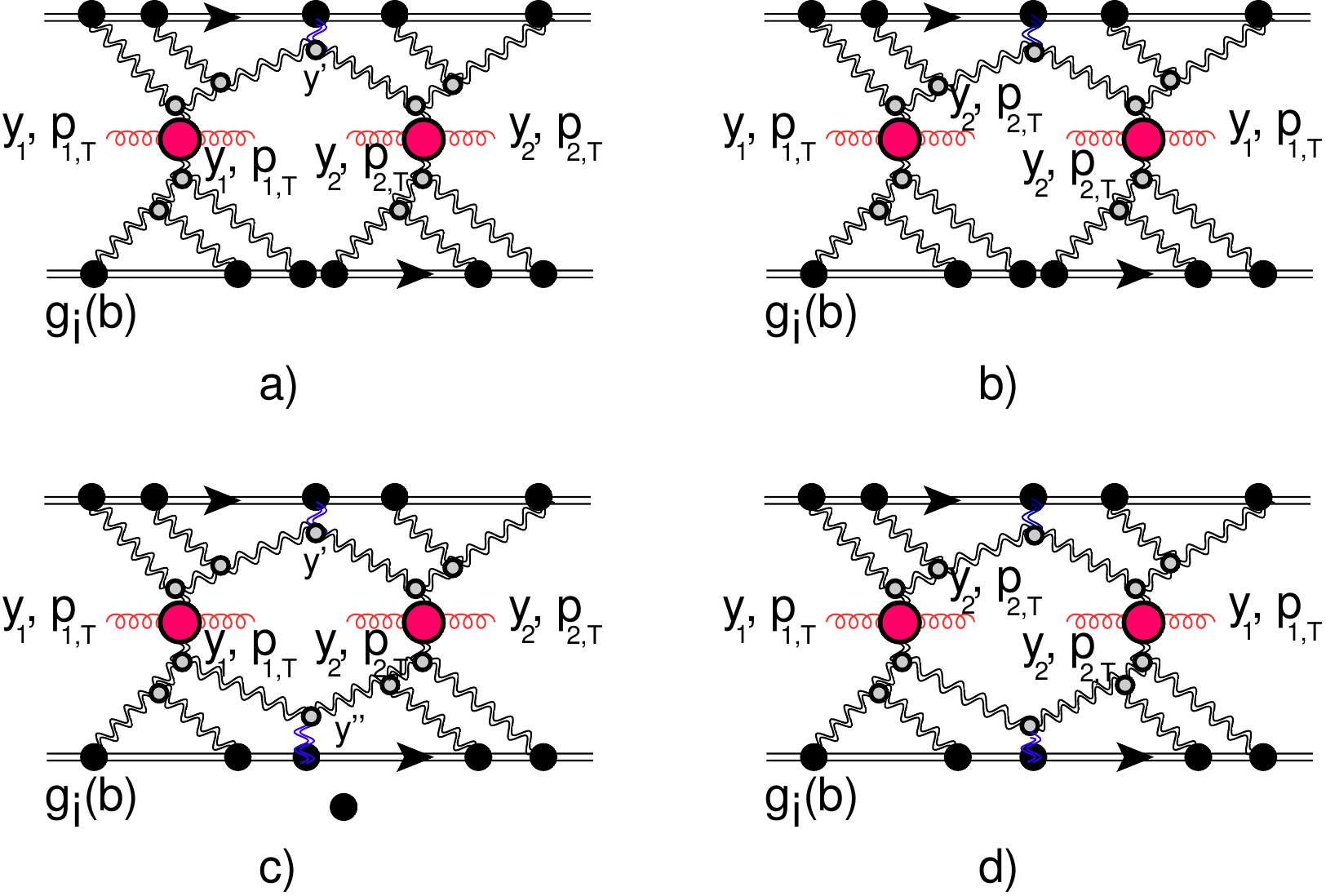}  
      \caption{Semi-enhanced and enhanced  diagrams:
 \protect\fig{se}-a and   \protect\fig{se}=c show  the cross sections 
 of double inclusive productions; \protect\fig{se}-b and
 \protect\fig{se}-d describe the interference diagram that
 leads to Bose-Einstein correlations.}
\label{se}
   \end{figure}

 \begin{boldmath}
  
  \subsection{Bose-Einstein correlation: contribution of the semi-enhanced
  and enhanced diagrams (diffraction production of large masses) }
   \end{boldmath}  
 In \fig{se} we show the diagrams in our model that have not been taken
 into account. They correspond to single diffraction in the region of large
 masses(\fig{se}-a and \fig{se}-b), and to double diffraction in two 
bunches
 of particles with large masses (\fig{se}-c and \fig{se}-d).   
 
 One can see from \fig{se}, that all these diagrams contain the 
integration
 over $y'$ . This integration is concentrated in the region $Y - y' 
\propto
 1/\Delta_{\rm BFKL}$, where $\Delta_{\rm BFKL} $ is the intercept of the
 BFKL Pomeron.  Preforming this integration, we reduce the diagrams of 
the upper
 part of \fig{se} to almost the same expression as it was used in the
 previous section, but instead of $g_i\Lb b \Rb$ we need to  insert 
the
 $b$- dependence of the triple Pomeron vertex, which in our model has the 
following
 form:
 \beq \label{SE1}
 \Gamma_{3 \pom}\,\,\propto\,\,e^{ - 2 m b}
 \eeq
 Bearing this in mind we can re-write $ \frac{d \sigma}{d y \,d^2 p_{T}
\,d^2 B\,d^2 b }$ of \eq{INC3}   in the form
 \bea \label{SE2}
&& \frac{d \sigma_{i,j}}{d y \,d^2 p_{T}\,d^2 B\,d^2 b }\,\,= \\
&& \frac{e^{-\,m\,b}}{S\Lb b, m_i\Rb}\,\frac{m^2 }{2 \,\pi\,\sqrt{g_i(0)\,\lambda}}\,\frac{2C_F}{\alpha_s (2\pi)^4}\,\frac{1}{p^2_T}\int  \,d^2 r \,e^{i \vec{p}_T\cdot \vec{r}}\,\,\nabla^2_T\,N^{i}_G\Lb Y - y; r; b \Rb\,\,\nabla^2_T\,N^{j}_G\Lb y; r; |\vec b-\vec B| \Rb\nn
\eea
In \eq{SE2} we restrict ourselves, by accounting only for interaction with
 the state $|1>$, as $g_1(0) \,\gg\,g_2(0)$.  Since in our 
model
we have  $m \,\gg\,m_1$, we can put $b=0$.  and reduce \eq{SE2} to 
\bea \label{SE3}
&& \frac{d \sigma_{i,j}}{d y \,d^2 p_{T}\,d^2 B\,d^2 b }\,\,= \\
&& \frac{e^{-\,m\,b}}{S\Lb b, m_i\Rb}\,\frac{m^2 }{2 \,\pi\,\sqrt{g_i(0)\,\lambda}}\,\frac{2C_F}{\alpha_s (2\pi)^4}\,\frac{1}{p^2_T}\int  \,d^2 r \,e^{i \vec{p}_T\cdot \vec{r}}\,\,\nabla^2_T\,N^{i}_G\Lb Y - y; r; 0 \Rb\,\,\nabla^2_T\,N^{j}_G\Lb y; r; |\vec B| \Rb\nn
\eea
for diagrams of \fig{se}-a and \fig{se}-b.

For the diagrams of \fig{se} -c and \fig{se}-d which correspond
 to double diffraction in large masses, we obtain
\bea \label{SE4}
&& \frac{d \sigma_{i,j}}{d y \,d^2 p_{T}\,d^2 B\,d^2 b }\,\,= \\
&& \frac{e^{-\,m\,b - m B}}{S^2\Lb b, m_i\Rb}\,\frac{m }{2 \,\pi\,g_i(0)\,\lambda}\,\frac{2C_F}{\alpha_s (2\pi)^4}\,\frac{1}{p^2_T}\int  \,d^2 r \,e^{i \vec{p}_T\cdot \vec{r}}\,\,\nabla^2_T\,N^{i}_G\Lb Y - y; r; 0 \Rb\,\,\nabla^2_T\,N^{j}_G\Lb y; r; 0 \Rb\nn
\eea
     
\begin{figure}[ht]
    \centering
  \leavevmode
  \begin{tabular}{c c c}
     \includegraphics[width=7cm]{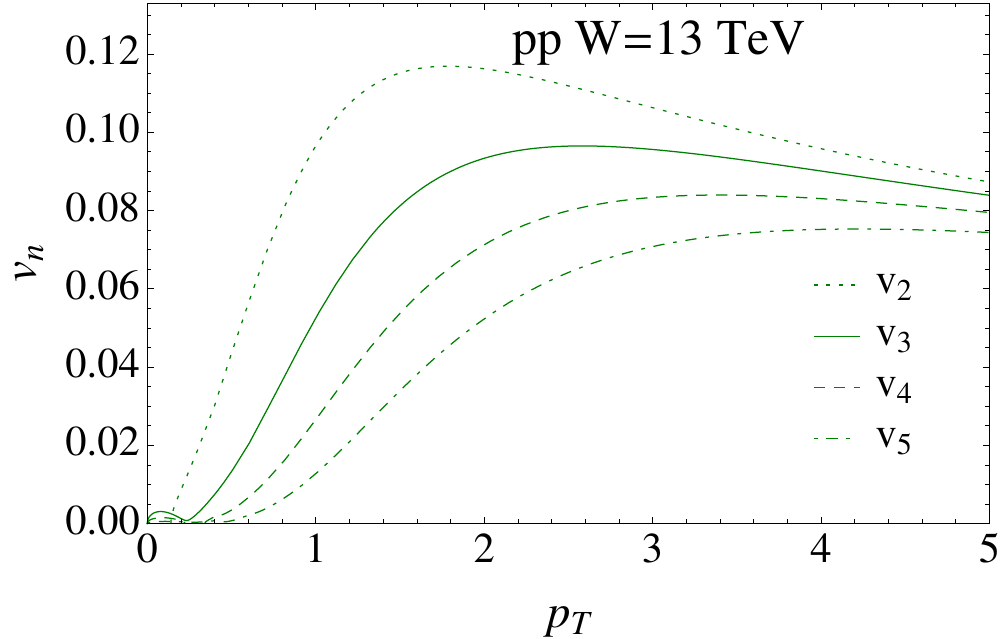}&~~~~~&
           \includegraphics[width=7cm]{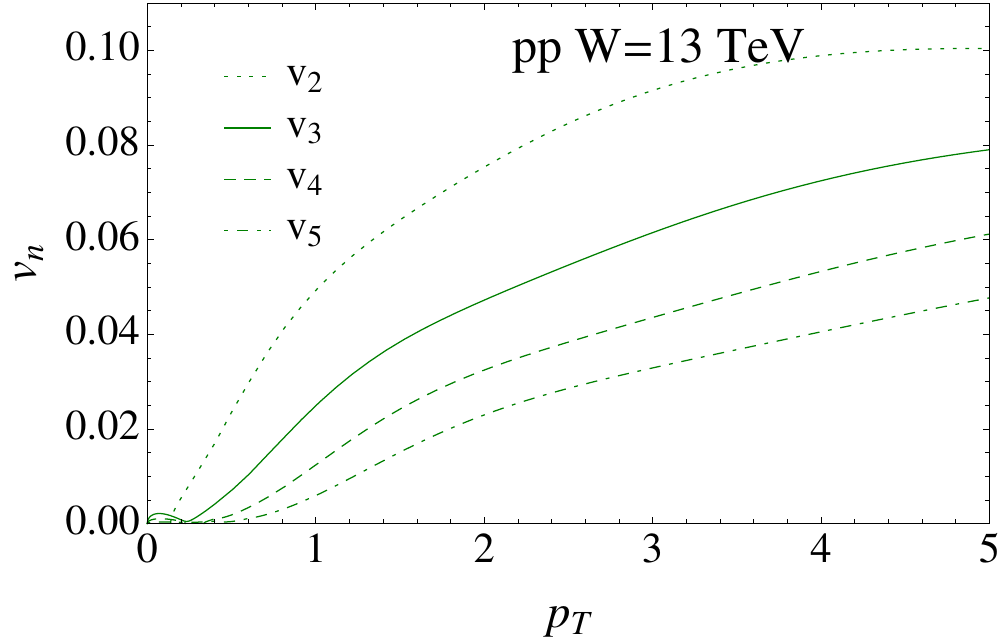}  \\
           \fig{ses}-a &  & \fig{ses}-b\\
           \end{tabular}
      \caption{$v_n$ versus $p_T$ at $W = 13\,TeV$ for  non-enhanced
 diagram of \protect\fig{incl} and   sum of all contributions.}
\label{ses}
   \end{figure}

Plugging  \eq{SE3} and \eq{SE4} into \eq{DINC11}, we can calculate the
 double inclusive cross sections. Using them and plugging in \eq{BEE3}
 and \eq{C}, we  obtain the correlation function and $v_n$. The result
 of these calculations is shown in \fig{ses}. One can see that 
contributions 
of semi-enhanced and enhanced diagrams, which are closely related to the
 processes of diffractive production of large masses in single diffraction
 (LMD-SD) and in double diffraction (LSD-DD), increase   the typical
 transverse momentum in $v_n$ dependence, on transverse momenta.

     
\begin{figure}[ht]
    \centering
  \leavevmode
     \includegraphics[width=10cm]{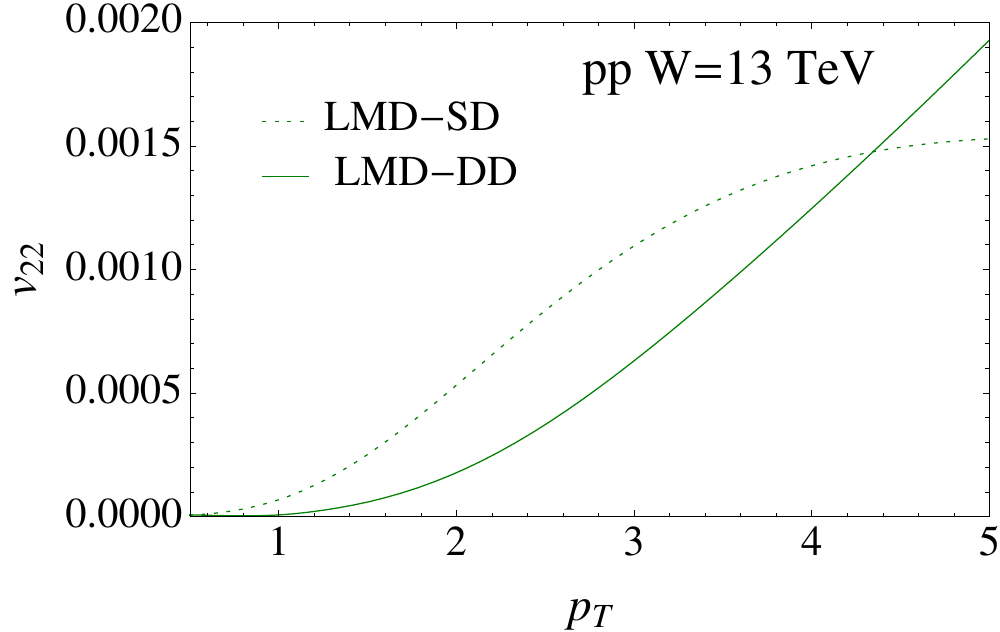}
      \caption{ The contribution to $v_{22}$ versus $p_T$ at $W = 13\,TeV$
 for   large mass diffraction in single(LMD-SD) and in double (LMD-DD).}
\label{qsen}
   \end{figure}

 Such behaviour is a direct consequence of the fact that typical momenta
 in the LSD contribution are of the order of $Q_s$, which is larger  than 
$m_1$
 and $m_2$, which determine  the hadron structure (see \fig{qsen})
     
Comparing \fig{ses}-a and \fig{ses}-b, shows that the typical momentum for
 the sum of the diagrams, is larger than for the non-enhanced diagrams.
   \fig{qsen} displays the dependence of $v_{n,n}$ in the semi-enhanced
 and enhanced diagrams. Comparing this figure with \fig{ses}-b, we note
 that the contribution of these diagrams are larger than the non-enhanced
 one, leading to the explanation of $p_T$ dependence in the experimental
 data of \fig{vnexp}. Therefore,   in our model the typical momentum is
  close to $Q_s$.

 \subsection{Comparison  with the experiment}
\begin{figure}[htbp]
  \centering
  \begin{minipage}[b]{0.45\textwidth}
    \includegraphics[width=\textwidth]{fig_11.pdf}
    \caption{Experimental data for $v_{nn}$  and $v_n$ versus $p_T$ at $W =
 13\,TeV$.\protect\cite{ATLASPP} }
    \label{vnexp}
  \end{minipage}
  \hspace{0.8cm}
  \begin{minipage}[b]{0.42\textwidth}
    \begin{tabular}{c c }
  \includegraphics[width= 0.5 \textwidth ]{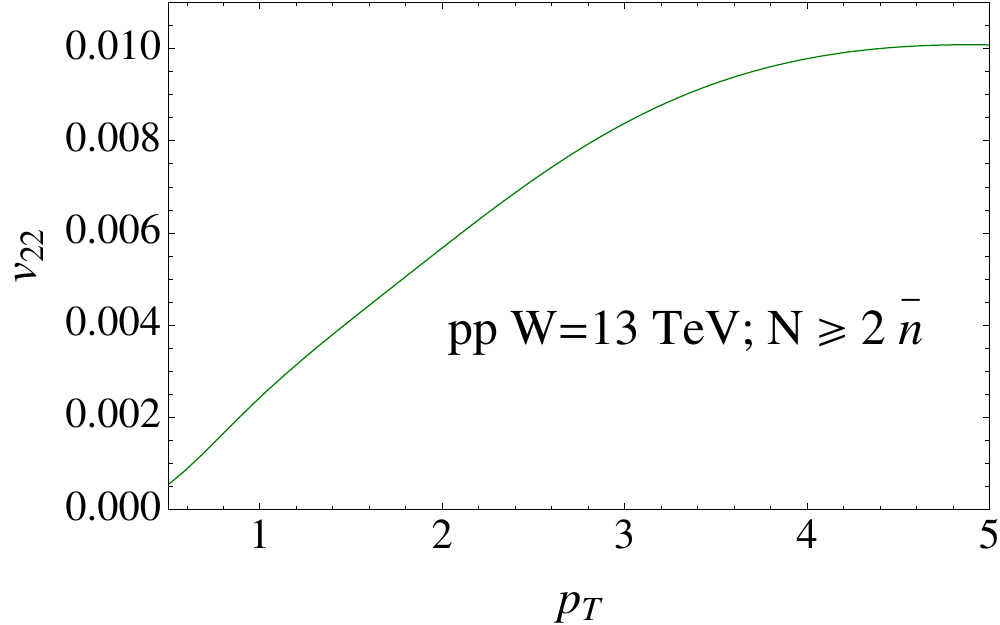}&   \includegraphics[width= 0.5 \textwidth ]{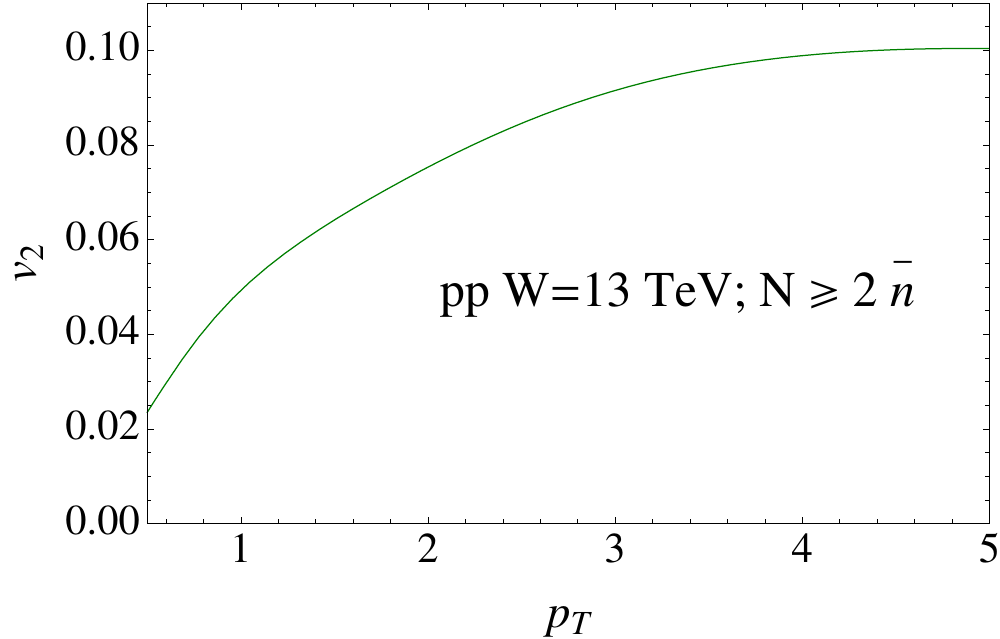}\\
   \includegraphics[width= 0.5 \textwidth ]{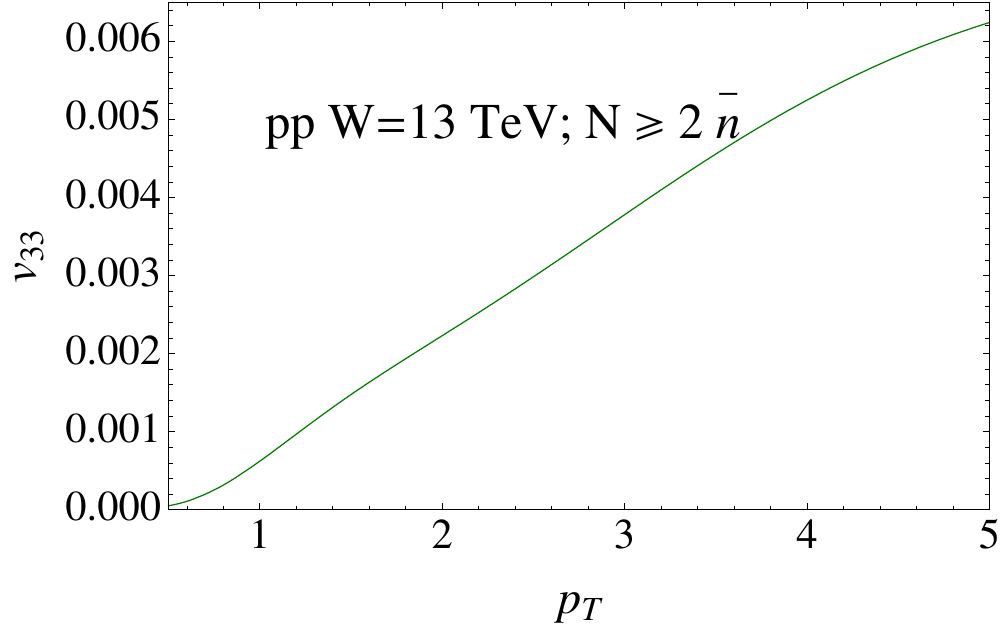}&   \includegraphics[width= 0.5 \textwidth ]{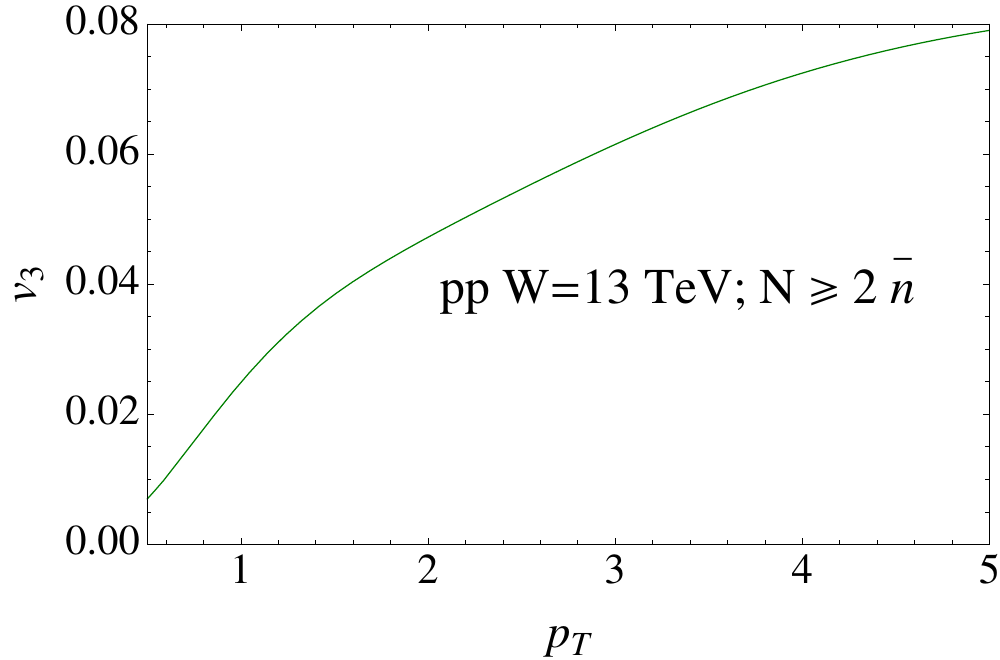}\\\end{tabular}    
    \caption{Our model for $v_{nn}$  and $v_n$ versus $p_T$  at $W=13\,TeV$.}
    \label{vnvspt}
  \end{minipage}
\end{figure}

.
  In \fig{vnexp} and  \fig{vnvspt} we plot the experimental
 data \cite{ATLASPP} and the results of our calculations.
 One can see that we predict the values and $p_T$ dependence
 of $v_n$ which are in  agreement with the experimental data.
 We  wish to stress that we used \eq{BEN1}-1 for the estimates
 of the values of $v_{n}$, but one can see that our prediction
 for $v_{n,n}$ are also in accord with the data. As we have mentioned
 the semi-enhanced and enhanced diagrams are closely related to
 the processes of large mass diffraction.  On the other hand,
 these processes give only 
   about  30\% contributions (see  Table 2). Indeed,  at $W =
 13\,TeV$ $R^{\rm lmd}_{\rm sd} = \sigma ^{\rm lmd}_{\rm sd}/(\sigma_{\rm el}
 +   \sigma ^{\rm smd}_{\rm sd} +  \sigma ^{\rm smd}_{\rm dd} )\,=\,0.26$ and
  $R^{\rm lmd}_{\rm dd} = \sigma ^{\rm lmd}_{\rm sd}/(\sigma_{\rm el} +  
 \sigma ^{\rm smd}_{\rm sd} +  \sigma ^{\rm smd}_{\rm dd})  \,=\,0.16$.    
   
   Such essential difference stems from the fact that the cross sections
 of diffractive production should be multiplied 	by the survival
 probability factor $\exp\Lb - 2 \Omega(r, Y - Y_0, b\Rb$ (see \eq{OMEGA}
 and Ref.\cite{GLM2CH}). This factor results in substantial suppression of
 the diffractive production, however, it is absent in the double inclusive 
cross
 sections.

 \section{Conclusions}
 
   In this paper we generalized  our model to include the hard 
processes and
 presented our estimates for $v_n$ for proton-proton collisions at
 high energy.  Our main result can be formulated  shortly:
  the model predicts  Bose-Einstein correlations which lead to
 the values of $v_n$,  that are  in accord with the experimental
 values.  Our estimates are obtained from a model which is  able to 
describe
 the typical soft observables for diffractive production, such as  total
 and elastic cross section and cross section of diffraction production, 
inclusive cross sections, long range rapidity correlations  and the deep
 inelastic $F_2$ structure function. In spite of being a phenomenological
 model which  parameterizes the data rather  than gives a theoretical
 interpretation, we believe that our model leads to reliable predictions
 for $v_n$ at high energies.   This belief is based  not only on the fact
 that the model describes both diffractive processes and processes of the
 multi-particle generation, but also  on the fact that it  includes 
all
 that we know from CGC on the behavior of the scattering amplitude in the 
saturation region. We showed that the angular correlations do not depend 
on energy
 and multiplicity, in accord with the experimental data.
  \par  Therefore,  before  making extreme assumptions on 
proton-proton
 collisions, such as the production of quark-gluon plasma in the large 
multiplicity
 events, we need to explain what  happens to  the Bose-Einstein 
correlations 
which are so large, that they are able to describe the angular
 correlations in the proton-proton scattering, without taking into
 account  interactions in the final state.


  {\bf Acknowledgements} 
   We thank our colleagues at Tel Aviv University and UTFSM for
 encouraging discussions. Our special thanks go to    
Carlos Cantreras, Alex Kovner and Michel  Lublinsky for
 elucidating discussions on the
 subject of this paper. 
 
 This research was supported by the BSF grant   2012124, by 
   Proyecto Basal FB 0821(Chile) ,  Fondecyt (Chile) grant  
 1140842, and by   CONICYT grant PIA ACT1406.  
 ~

 ~

      \end{document}